\documentclass[12pt]{iopart}
\eqnobysec
\usepackage{tabularx}
\usepackage{graphicx}
\usepackage{bm}
\usepackage{exscale}
\usepackage{amssymb}
\usepackage{amsfonts}
\usepackage{amsbsy}

\def\text#1{\mbox{\rm #1}}
\def\eqref#1{(\ref{#1})}

\def\beq{\begin{equation}}
\def\eeq{\end{equation}}
\def\beqn{\begin{eqnarray}}
\def\eeqn{\end{eqnarray}}

\def\C60{A$_x$C$_{60}$}

\def\SROone{ Sr$_{2}$RuO$_{4}$}

\def\HgCu3{HgCa$_2$Cu$_3$O$_{8+y}$}
\def\HgCu4{HgBa$_2$Ca$_3$Cu$_4$O$_{10+y}$}
\def\TlCu{Tl$_2$Ba$_2$CuO$_{6+\delta}$}
\def\TlCu3{Tl$_2$Ba$_2$Ca$_2$Cu$_3$O$_{10+y}$}
\def\TlCu4{Tl$_2$Ba$_2$Ca$_3$Cu$_4$O$_{12+y}$}

\def\BiCu3{Bi$_2$Sr$_2$Ca$_{2}$Cu$_3$O$_y$}

\def\8LSCO{La$_{1.88}$Sr$_{.12}$CuO$_4$}
\def\110LNSCO{La$_{1.5}$Nd$_{0.4}$Sr$_{0.1}$CuO$_{4}$}
\def\stage4LCO{La$_{2}$CuO$_{4+\delta}$}
\def\Y248{YBa$_2$Cu$_4$O$_8$}

\def\NbSe2{NbSe$_2$}
\def\TaSe2{TaSe$_2$}
\def\TiSe2{TiSe$_2$}
\def\NaCoOH2O{Na$_{0.3}$CoO$_{2y}$H$_2$O}
\def\MgB2{MgB${}_2$}

\def\tr{\mbox{tr}}

\def\bra#1{\langle#1\vert}
\def\ket#1{\vert#1\rangle}
\def\ipr#1#2{\langle#1\vert#2\rangle}
\def\me#1#2#3{{\langle#1\vert#2\vert#3\rangle}}

\def\s12{spin-${\frac{1}{2}}$}

\begin{document}

\title[Entanglement Scaling at Quantum Lifshitz fixed points and topological fluids]{Scaling of Entanglement Entropy at 2D quantum Lifshitz fixed points and topological fluids}
\author{Eduardo Fradkin}

\address{Department of Physics, University
of Illinois at Urbana-Champaign, 1110 West Green Street, Urbana, Illinois 61801-3080, USA}

\date{\today }

\begin{abstract}
The entanglement entropy of a pure quantum state of a bipartite system is
defined as the von Neumann entropy of the reduced density matrix obtained by tracing
over one of the two parts. Critical ground states of local Hamiltonians in one dimension
have an entanglement entropy that diverges logarithmically in the subsystem size, with a universal
coefficient that is is related to the central charge of the associated
conformal field theory. Here I will discuss recent extensions of these ideas to a class of quantum critical points with dynamic critical exponent $z=2$ in two space dimensions and  to 2D systems in a topological phase.  ÊThe application of these ideas to quantum dimer models and fractional quantum Hall states will be discussed.
\end{abstract}


\maketitle

\setcounter{footnote}{0}
\setcounter{section}{0}

\section[Introduction]{Introduction}
\label{sec:introduction}

The entanglement of macroscopic systems measures non-local correlations of a uniquely quantum mechanical nature. Quantum mechanical {\em entanglement} is defined in terms of observing the state of a (large) subsystem of a much larger system, which will is assumed to be in a pure quantum state. The {\em entanglement entropy} is a quantitative measure of these non-local quantum correlations.

Let $\Omega$ be a macroscopic quantum system, {\it i.e.\/} a system with a thermodynamically large number of degrees of freedom, and $A$ be a (macroscopically large) subsystem, $A \in \Omega$. Let $B$ be the {\em complement} of $A$, such that $A$ and $B$ constitute a {\em partition} of $\Omega$: $A \bigcup B=\Omega$ with $A \bigcap B=\emptyset$, and $\Gamma$ their common boundary. Let us consider a {\em pure quantum state} $\ket{\Psi}$ (suitably normalized $\ipr{\Psi}{\Psi}=1$) of the macroscopic system defined on the whole region $\Omega$. For this pure state, the reduced density matrix for region $A$, $\rho_A$, and region $B$, $\rho_B$, are the Hermitian operators
\begin{equation}
\rho_A={\tr}_B \ket{\Psi}\bra{\Psi}, \qquad \rho_B={\tr}_A \ket{\Psi}\bra{\Psi}
\label{eq:reduced-dms}
\end{equation}
where $\tr_A$ and $\tr_B$ denote the trace over the degrees of freedom on $A$ and $B$, respectively.  The reduced density matrices $\rho_A$ and $\rho_B$  obey the obvious normalization conditions:
\begin{equation}
\tr_A \rho_A=\tr_B \rho_B=1
\label{eq:normalization}
\end{equation}
Under these circumstances, th subsystems $A$ and $B$ are in a {\em mixed quantum state} defined by their reduced density matrices.

From their definition, it follows that the density matrices are Hermitian operators whose eigenvalue spectra, $\{\lambda_n^A\}$ and $\{ \lambda_n^B\}$ (respectively), are {\em positive} and lie in the range, $0\leq \lambda_n^A, \lambda_n^B \leq 1$. Here $n$ is a complete set of labels of the spectrum of eigenstates of each reduced density matrix. Thus, the eigenvalues are the {\em probabilities} of finding region $A$ in a given eigenstate of its reduced density matrix $\rho_A$.

With the above definitions, the von Neumann entanglement entropy $S_A$ for {\em observing} region $A$, is given by
\begin{equation}
S_A=-\tr_A \left(\rho_A \ln \rho_A\right)=\sum_{n} \lambda_n^A \ln \lambda_n^A
\label{eq:von-neumann}
\end{equation}
and the same for subsystem $B$. The von Neumann entropy
has the important symmetry property: $S_A=S_B$, which holds provided $\Omega$ is in pure state $\ket{\Psi}$.
On the other hand, if the two subsystems $A$ and $B$ are physically separate, the wave function factorizes into two (pure) states with support in $A$ and $B$ respectively: $\ket{\Psi}=\ket{\Psi_A} \otimes \ket{\Psi_B}$. In this case, and only in this case, the entanglement entropy vanishes, $S_A=S_B=0$, as each subsystem is in a precisely known pure state.

The standard notions discussed above are simple to realize in the context of small quantum mechanical systems. Thus, for a simple system of just two spin $1/2$ degrees of freedom in the singlet state $\ket{\Psi}=(1/\sqrt{2}) \left(\ket{\uparrow, \downarrow}- \ket{\downarrow, \uparrow}\right)$, the {\em probability} of measuring the left spin in state $\ket{\uparrow}$($\ket{\downarrow}$) is $1/2$ ($1/2$). For this trivial case the entanglement entropy of the left spin (part of this singlet state)  is (quite obviously) $S=\ln 2$.

However, in the case of macroscopic systems as well as for quantum field theories, the general behavior of the entanglement entropy is far less obvious and, except in very simple cases, has a complex behavior. It is a non-local property of the wave function as a whole of a system with an infinite number of degrees of freedom.  There are several reasons for being interested in such a non-local property of the wave function. In principle, in a local field theory, one expects that the behavior of local operators should yield a complete picture of its possible behaviors. However, the wave function of a quantum mechanical system (including a quantum field theory) contains non-local behaviors which are encoded in the entanglement. Such non-local information underlies the degree of difficulty of computing the wave function of an extended system when starting from some simple ansatz usually based on simple local properties.

The entanglement entropy in principle depends not only on the intrinsic parameters of the theory, encoded in the local Lagrangian, but also on the size and shape of the region being observed.
Let us consider the problem of the {\em scaling of  quantum entanglement} as the region that is being observed becomes macroscopically large. Let $L$ be the linear size of the {\em entire} system, and $L_A$ the linear size of the observed subsystem ($A$), where we will be interested in the regime $L \gg L_A\gg a$, where $a$ is a short distance scale (the ``UV cutoff''). It has been know for quite a long time\cite{bombelli-1986,srednicki-1993} that free massive field relativistic theories in $d+1$ space-time dimensions obey a so-called {\em area law}, {\it i.e.\/} a scaling of the entanglement entropy of the form $S_A= \textrm{const}.\, \left(L_A/a\right)^{d-1}$, which scales with the {\em area} of the boundary of region $A$, instead of a {\em volume} scaling as with the thermodynamic entropy. Indeed, the main motivation of the early interest in the area law was the formal analogy between the area law behavior of the entanglement entropy and the Bekenstein-Hawking entropy of a black hole\cite{bekenstein-1973,hawking-1975} $S_{BH}=k_B A/4\ell_P^2$, where $k_B$ is the Boltzmann constant,  $A$ is the area of the event horizon of the black hole and $\ell_P=\sqrt{G\hbar/c^3}$ is the Planck length ($G$ being Newton's universal gravitation constant). However, while in the General Relativity context the prefactor of the  Area Law is universal, in the case of the entanglement entropy is not as it depends on the choice of the short distance cutoff $a$. At any rate, the area law of the scaling of the entanglement entropy holds for any local field theory.\footnote{Fermi systems at {\em finite density} (Landau Fermi liquids) are known to exhibit logarithmic violations the area law, $L_A^{d-1} \ln L_A$.\cite{wolf-2006,klich-2006b} This behavior is due to the existence of a Fermi surface and its quantum fluctuations.}

In this paper we will discuss the problem of universality in the scaling behavior of the entanglement entropy. Having noted above that, in general, for a local quantum field theory the leading behavior is an area law with a non-universal coefficient,  we will be interested on whether the entanglement entropy may have universal {\em sub-leading} terms, {\it i.e.\/} universal corrections to the area law. For a generic local theory universality is not expected unless the field theory itself is scale invariant. In other terms, the quantum field theory must be a quantum critical system, {\it i.e.\/} a quantum field theory at a renormalization group fixed point describing a continuous quantum phase transition.

The best and most completely understood quantum field theories at a critical point are conformal field theories (CFT) in $1+1$ dimensions. In this case, the area law itself is replaced by a logarithmic dependence on the linear size of the region, $S=\frac{c}{3} \ln (L_A/a)$, whose prefactor is indeed universal, and it is given in terms of the central charge $c$ of the CFT \cite{Callan1994,Holzhey-1994,Calabrese2004,Vidal2003}. In general space-time dimension, scale-invariant quantum field theories are RG fixed points. At a fixed point the RG predicts the scaling behavior of {\em local} operators. In practice, again with the exception of CFTs in $1+1$ dimensions, such fixed point theories are generally understood in terms of various perturbative RGs, namely the $4-\epsilon$ expansion, the $2+\epsilon$ expansion, and the $1/N$ expansion (see, {\it e.g.\/} Ref\cite{zinn-book,cardy-book}). 

In this language there is little conceptual difference between a fixed point theory describing a classical thermodynamic phase transition and a fixed point of a quantum field theory. Indeed, in the case of relativistic quantum field theories they are related by an analytic continuation from Minkowski to Euclidean space-times. In essence, of the entire conceptual construction of the theory of quantum critical systems is based on this formal connection. However, in general,  relativistic invariance may or may not be present and the theory has a dynamic critical exponent $z$, relating space and time, plus a set of scaling laws that describe the effects of thermal fluctuations (in effect, a form of finite size scaling) (see {\it e.g.\/} Ref. \cite{sachdev-book}). 

The description of scale-invariant quantum field theories is the problem of the scaling behavior of {\em local} observables. Little is known in general of the behavior of non-local observables such as the entanglement entropy in quantum field theories.\footnote{For a recent review of entanglement in  quantum many-body systems see Ref.\cite{Amico2008}.}
 However, one may ask if the scaling   behavior of quantum entanglement is related in any simple way to the fixed point theory of local observables and, in particular, what is the theory of the scaling of quantum entanglement. In the case of CFTs in $1+1$ dimensions this problem is by now reasonably understood, and the connection between entanglement entropies and the structure of the CFT is known. (For a comprehensive discussion see Ref.\cite{calabrese-2009b}).\footnote{Logarithmic scaling of the entanglement entropy was also found at infinite disorder fixed point of random quantum Heisenberg chains\cite{Refael2004,Refael2007}. These systems that are scale invariant only in the sense of an ensemble average, and the physical meaning of the universal coefficient of the entanglement entropy is presently not understood.}
 
 In this paper we will discuss the problem of universality in the scaling of the entanglement entropy in the context of a special class of quantum critical points in $2+1$ dimensions with dynamic critical exponent $z=2$: the conformal quantum critical points of the quantum Lifshitz universality class(es) introduced in Ref.\cite{Ardonne2004}. These fixed-point theories in $2+1$ dimensions are interesting in that they are essentially solvable. Indeed, as shown by Ardonne et al\cite{Ardonne2004}, these theories have the special feature the the ground state (or ``vacuum'') wave function is {\em scale invariant}. In contrast, in most scale-invariant theories (including CFTs in $1+1$-dimensions) it is the {\em action} (and the associated path-integral) that is scale (and conformal) invariant but the wave function {\em scales}. The scale (and conformal) invariance of the quantum Lifshitz wave functions (and of their norms) is the reason for the solvability of these quantum critical points. In particular, the weight of a field configuration of a wave function of the quantum Lifshitz class has the same form as the Gibbs weight of a system in two-dimensional classical statistical mechanics which is scale invariant as it is a classical critical point. This 2D quantum - 2D classical connection has been exploited with great success in quantum dimer models at the Rokhsar-Kivelson point\cite{Rokhsar1988,Fradkin1991} and its many generalizations\cite{Fradkin1990,Moessner2001,Ardonne2004,Castelnovo2005,Fendley2005,Papanikolaou2007b}.  The quantum Lifshitz fixed points are multicritical points separating various ``valence bond'' ordered phases (that spontaneously break spatial symmetries of the underlying lattice model) and are also proximate to topological phases. For this reason these fixed point theories  typically have more than one relevant operator. In a number of cases this leads to the transition to become first order while in others it remains continuous.\cite{Fradkin2004,vishwanath2004}. They are also intimately related to $\mathbb{Z}_2$ gauge theories and, hence, their relevance for the description of topological (deconfined) phases.\cite{Fradkin1990,read1991,Moessner2001,Moessner2002}
 
 The conformal structure, and their locality, of the ground state wave functions of quantum Lifshitz fixed points allows for a direct computation of the entanglement entropy and to investigate its scaling properties for this class of states\cite{Fradkin2006,Papanikolaou2007,hsu-2009}. Indeed, the entanglement entropy of the 2D wave functions is simply related to a combination of free energies of the related 2D Euclidean CFT obeying specific boundary conditions, whose scaling with size is well known\cite{Cardy1988}. In Refs.\cite{Fradkin2006,hsu-2009} it was found that  the entanglement entropy at a general quantum Lifshitz fixed point scales as \footnote{For specific geometries, {\it i.e.\/} a disk, there are also universal finite contributions depending on the aspect ratio $\frac{L_A}{L_B}$ of the chosen geometry\cite{Fradkin2006,hsu-2009}.}
 \begin{equation}
 S_A=\alpha \, \left(\frac{L_A}{a}\right) + \gamma_{QCP} + O\left(\left(\frac{a}{L_A}\right)\right)
\label{eq:entropy-q-lifshitz}
 \end{equation}
provided the boundary $\Gamma$ of region $A$ is smooth and has no cusps (or other singular curvatures).
Here $\alpha$, the prefactor of the ``area'' law (a perimeter law in this 2D case) is once again non-universal, and $\gamma_{QCP}$ is a universal {\em finite} contribution  to the entanglement entropy determined by the conformal structure of the wave function\cite{hsu-2009}. Eq.\eqref{eq:entropy-q-lifshitz} yields the first result on the universal scaling of the entanglement entropy for space dimensions $d>1$.

In contrast, the scaling of the entanglement entropy at relativistic fixed points in space dimensions $d>1$ is much less understood and has only begun to be considered quite recently. A number of important first results are given in Ref.\cite{Metlitski-2009} for the relativistic $\phi^4$ field theory with a global $O(N)$ symmetry, whose entanglement entropy was found to obey the same scaling found at the quantum Lifshitz fixed points, {\em c.f.\/} Eq.\eqref{eq:entropy-q-lifshitz}. The structural origin of the universal coefficients in these more generic fixed point theories is not yet well understood.
 
The other class of macroscopic physical systems for which the scaling of entanglement is understood are topological phases of matter in two space dimensions and their associated topological quantum field theories, where it is found that they obey the scaling\cite{Kitaev2006a,Levin2006} 
\begin{equation}
S_A=\alpha \, \left(\frac{L_A}{a}\right) - \gamma_{\rm topo} + O\left(\left(\frac{a}{L_A}\right)\right)
\label{eq:entropy-topo}
 \end{equation}
where, once again, $\alpha$ is non-universal, and  the finite term, $-\gamma_{\rm topo}$ is universal and given in terms of the effective quantum dimension $\mathcal{D}\equiv\sqrt{\sum_a d_a^2}$ of the excitations of the topological phase; here $\{ d_a\} $ are the quantum dimensions of excitations of type $a$. Notice the opposite sign in Eqs.\eqref{eq:entropy-q-lifshitz} and \eqref{eq:entropy-topo}.

Being topological, these theories have no dependence on any length scale ({\it i.e.\/} the correlation length is zero). Indeed, these theories are not only scale invariant but they have a much larger symmetry as they are effectively independent of the metric of the two-dimensional space on which they exist. Thus, provided the short distance structure is properly treated, topological phases are well described by topological quantum field theories. The prototype topological phases are the two-dimensional electron gases (2DEG) in large magnetic fields in the fractional quantum Hall (FQH) regime\cite{Laughlin1983}, which behave as topological fluids\cite{Wen1990,Wen1995}. Their effective field theories are intimately related to Chern-Simons gauge theory\cite{Witten1989}. The other class of topological phases of matter (as yet not realized experimentally) are the deconfined phases of $\mathbb{Z}_2$ gauge theories and its generalizations\cite{Fradkin1979,Krauss1989,Dijkgraaf1990a,Preskill1990,Bais1992,Kitaev2003,Levin2005,Fendley2005,Fidkowski2009}.
 
We will see below that the entanglement entropy of a topological phase (and of a topological field theory) depends on the topology of the surface, on the state on the system on that surface,  and on the topology of the region being observed.\cite{Dong2008} It further depends on whether the state whose entanglement is being tested is one of the (generally degenerate) vacua of the topological phase, or if it is an excited state ({\it i.e.\/} if there are excitations carrying non-trivial labels). This feature of the entanglement entropy of a topological phase raises the question of whether there is a minimum number of entanglement entropy measurements that can uniquely determine the structure of the topological field theory. If this were true it would constitute a classification of topological field theories (at least in 2D). This is a well known open problem (see, Ref.\cite{moore-1989a,moore-1989b}).

In this paper we present a review of recent results on the scaling of the entanglement entropy at 2D quantum Lifshitz fixed points and in topological phases and field theories. We begin with a brief description of topological phases and their associated quantum field theories (Section \ref{sec:topological}). We then discuss  the computation and scaling of quantum entanglement entropy for the quantum Lifshitz universality class (Section \ref{sec:entanglement-quantum-lifshitz}). In Section \ref{sec:chern-simons} we compute the entanglement entropy for Chern-Simons gauge theory and for the associated FQH fluids. The conclusions are presented in Section \ref{sec:outlook}.

\setcounter{footnote}{0}
\section[Topological Phases]{Topological Phases of Matter of Topological Field Theory}
\label{sec:topological}

Topological phases of matter are liquid phases of electron fluids and spin systems without long range order, with or without  time reversal symmetry breaking. The quasiparticles of these fluid phases are vortices typically with fractional charge (if the fluid is charged as in the case of the FQH states of the 2DEG) and fractional statistics\cite{Wilczek1982} which in turn may be  Abelian and non-Abelian. 

Topological phases do not have local order parameters. Instead their observables are non-local objects similar to the Wilson and t'Hooft loops of gauge theories. This similarity is not accidental as the effective field theories of topological phases are topological gauge theories. Thus, they have a hidden topological order and a concomitant topological vacuum degeneracy which depends on the topology of the 2D surface on which the state is defined.\cite{Wen1990,wen-1998} 

The wave functions of the quasiparticles associated with topological phases are states that transform non-trivially under a braiding operation\cite{arovas-1984} and hence exhibit fractional statistics. More abstractly, the quasiparticles  transform like irreducible representations of the Braid group. (For a good modern review see J. Preskill\cite{Preskill2004}.)  In most cases, {\it e.g.\/} as in the Laughlin FQH states and their generalizations, these representations are one-dimensional and classified in terms of a single number, the statistical phase. There are a number of systems, notably the non-Abelian FQH states\cite{moore-read-1991,nayak-wilczek-1996,Read-Rezayi-1999,fradkin-nayak-schoutens-1999} in which the quasi-particles are in non-Abelian (finite-dimensional) representations of the Braid group. In this context, this means that the states with a fixed number of quasiparticles are not completely specified by giving just their coordinates. In fact there are a finite number of linearly independent states for a fixed set of quasiparticle coordinates. This defines a topologically protected finite-dimensional quasiparticle Hilbert space. This feature of the non-Abelian FQH states has been the primary motivation behind the concept of {\em topological quantum computing}\cite{Freedman2000,Freedman2002,Kitaev2003}  and it is a subject of intense research\cite{dassarma-2005,dassarma-2007}.
 
At very low energies, {\it i.e.\/} at energies low compared to the quasiparticle gap, topological phases admit an effective field theory description: Topological Field Theory, \emph{e.g.}, Chern-Simons gauge theory, discrete gauge theory. The best known examples are the fractional quantum Hall fluids (described by Chern-Simons gauge theory) and $\mathbb{Z}_2$ deconfined phases, which describe quantum dimer models\cite{Moessner2001} and Kitaev's Toric Code\cite{Kitaev2003}.

Experimentally, the ``best  known''  topological quantum liquids are, as we stated above, the FQH states of the  2DEG at large magnetic fields. Most of the FQH fluids are Abelian FQH states: Laughlin\cite{Laughlin1983} and their generalization, the Jain states\cite{jain-1989}. These states have been know to exist as plateaus in the Hall conductance: $\sigma_{xy}=\nu e^2/h$ where $\nu$ is the filling fraction of the partially filled Landau level. As far back as 1998 it was determined experimentally, through measurements of the noise of the tunneling current at a constriction of the FQH fluid, that their excitations are fractionally charged.\cite{dePicciotto-1997,Samindayar-1997}. More recent experiments\cite{Camino-2005} have attempted to measure the fractional statistics of the (Abelian) quasiparticles by means of a quantum interferometer along the lines proposed theoretically earlier on by Chamon et al\cite{chamon-1997}.

More recent experimental work has focused on the non-Abelian FQH states. The best and most promising candidate is the FQH state in the first Landau level, at filling fraction $\nu=5/2$. This state has been suspected to be a Pfaffian (Moore-Read) FQH state (firm candidate). There is now  strong evidence for $q=e/4$ vortex at $\nu=5/2$ from
 shot noise at a point contact experiments\cite{dolev-2008},
and DC transport experiments also at a point contact \cite{radu-2008}.
There is also a well defined plateau in the Hall conductance at $\nu=12/5$, and a possible candidate is a parafermion state (Read-Rezayi) state.\footnote{An Abelian (Jain) $2+2/5$ state is however a possible competitor.}

It has also been suggested\cite{cooper-2001} that rapidly rotating ultra-cold Bose gases may also be in possible non-Abelian (Pfaffian) FQH state of bosons at $\nu=1$ (a relative of the Moore-Read fermionic state\cite{moore-read-1991}) but the experiment is hard and difficult to do.
The recent discovery\cite{xia-2006,kidwingira-2006} of a superconducting state with apparently spontaneous time-reversal symmetry breaking in {\SROone} suggesting that it is a $p_x+ip_y$ superconductor has raised the possibility that this superconductor may have non-Abelian half-vortices\cite{ivanov-2001,stern-2004}.\footnote{The current experimental evidence is strong but not uncontroversial as the predicted edge states have not (yet) been observed.}

\subsection{Hydrodynamic Picture of Topological Phases and Gauge Theory}
\label{sec:hydrodynamic}

We will begin with the effective field theory of the Abelian fractional quantum hall states. The Abelian FQH states are generalizations of the Laughlin states whose wave functions for a system of $N$ electrons in $N_\phi$ magnetic fluxes, at filling fraction $\nu=N/N_\phi=1/m$ (with $m$ an odd integer) are\cite{Laughlin1983}
\begin{equation}
\Psi_m\left(z_1,\ldots,z_N\right)=\prod_{i<j} \left(z_i-z_j\right)^m \; e^{-\sum_i |z_i|^2/4\ell^2}
\label{eq:laughlin}
\end{equation}
where $z=x+iy$ are the complex coordinates of the plane and $\ell=\sqrt{\hbar c/eB}$ is the magnetic length. 

The effective field theory of the Laughlin FQH states is an Abelian Chern-Simons gauge theory\cite{zhang-1989,lopez-1991} and it is usually derived through a mean field theory approach (see, {\it e.g.\/} Ref.\cite{lopez-fradkin-1998}). We will follow here instead a more intuitive phenomenological approach of Refs.\cite{frohlich-kerler-1991,frohlich-zee-1991,Wen1995} based on hydrodynamic considerations. Here I follow in detail the approach described by Wen\cite{Wen1995}.

The 2DEG is a charged fluid. As such it has a conserved charge current $j_\mu$, $\mu=0,1,2$, using a 3-vector relativistic notation.
From the conservation of the current
\begin{equation}
\partial_\mu j^\mu=0 
\end{equation}
it follows that in 2D the current is {\em dual} to a vector field $\mathcal{A}_\mu$, defined by
\begin{equation}
j_\mu=\frac{1}{2\pi} \epsilon_{\mu \nu \lambda}\partial^\nu \mathcal{A}^\lambda 
\end{equation}
The vector field $\mathcal{A}_\mu$ is a {\em gauge field} since a gauge transformation $\mathcal{A}_\mu \to \mathcal{A}_\mu +\partial_\mu \Lambda$ {\em leaves the currents $j_\mu$ invariant}. 

Hence, we expect to be able to construct an effective theory of the FQH fluid solely in terms of the vector field $\mathcal{A}_\mu$ (hereafter referred to as the hydrodynamic gauge field\cite{Wen1995}). However, to construct this theory we must take into account that the FQH fluid is {\em incompressible}, {\it i.e.\/} that is a fully gapped state (except possible at the edges of the system where its chiral edge states reside\footnote{For an extensive discussion see Ref.\cite{Wen1995}.}). Since the theory is gapped and gauge-invariant we expect the effective action for the hydrodynamic gauge field to be gauge invariant (as in any gauge theory) and local. Consequently the action is a local gauge invariant function of $\mathcal{A}_\mu$. In addition the effective action must be odd under time reversal, since the 2DEG is in the presence of a large perpendicular magnetic field. 

These requirements fully determine the form of the effective action since the only possible form of the effective action for the hydrodynamic gauge field $\mathcal{A}_\mu$ that is local, gauge invariant and odd under time-reversal is the Chern-Simons action for a gauge field $\mathcal{A}_\mu$ with a $U(1)_m$ gauge invariance:
\begin{equation}
S(\mathcal{A})= \frac{m}{4\pi} \int_{\Sigma\times S^1} d^3x \; \epsilon_{\mu \nu \lambda} \mathcal{A}^\mu \partial^\nu \mathcal{A}^\lambda 
\end{equation}
Here $\Sigma$ is the spatial manifold and $S^1$ is the (compactified) time direction. Furthermore, 
the requirement that the action must also be invariant under large gauge transformation on a closed manifold $\Sigma$ (such as the sphere $S^2$ or a torus $T^2$) further restricts the theory by {\em quantizing} the parameter $m$ to be an {\em integer}\cite{Witten1989}.

The form of the effective theory is now fully determined once the coupling to an external electromagnetic field $A_\mu$ is specified (consistent with invariance under ordinary electromagnetic gauge transformations) with a term in the local Lagrangian of the form $\mathcal{L}_{\rm int}=-e j_\mu A^\mu = -\frac{e}{2\pi} \epsilon_{\mu \nu \lambda} A^\mu \partial^\nu \mathcal{A}^\lambda$, and that the quasiparticles (the vortices of the fluid) couple through their currents $j_\mu^{qp}$ to the hydrodynamic gauge field $\mathcal{A}_\mu$ with the standard gauge invariant minimal coupling. This theory then predicts\cite{Wen1995} that the Hall conductance is $\sigma_{xy}=\frac{1}{m} \frac{e^2}{h}$, and that its excitations are vortices with fractional charge $q=e/m$ and fractional (braid) statistics $\theta=\pi/m$. For a system of fermions the {\em level} $m$ of the Chern-Simons gauge theory must be an odd integer.

Following Wen again\cite{Wen1995}, we now note that a 2DEG on a manifold $\Sigma$ with a boundary $\partial \Sigma$ has chiral edge states, described by a compactified chiral boson CFT $U(1)_m$ with compactification radius $R=1/\sqrt{m}$ and central charge $c=1$.

With some work, the hydrodynamic description also generalizes to the non-Abelian FQH states\cite{fradkin-nayak-tsvelik-wilczek-1998,fradkin-nayak-schoutens-1999,fradkin-nayak-2009}.
I will not give a full description here as this will be discussed in upcoming publication\cite{fradkin-nayak-2009}. 
For our present purposes it will be sufficient to note the structure of the effective theory in a few cases of interest. 

The wave functions for the non-Abelian Moore-Read (MR) (Pfaffian) FQH states are
\begin{equation}
\Psi_q(z_1,\ldots,z_N)=\textrm{Pf}\left(\frac{1}{z_i-z_j}\right)\; \prod_{i<j} \left(z_i-z_j\right)^q \; e^{-\sum_i |z_i|^2/4\ell^2}
\label{eq:MRq}
\end{equation},
where Pf denotes the pfaffian of the antisymmetric matrix $1/(z_i-z_j)$.

For the $\nu=1$, the wave function is  a  MR bosonic state with $q=1$. It turns out\cite{fradkin-nayak-tsvelik-wilczek-1998,fradkin-nayak-schoutens-1999} that in this case the effective field theory is an $SU(2)_2$ Chern-Simons theory. 
On the other hand, for the case of the $\nu=5/2$ fermionic pfaffian and anti-pfaffian states, the theory now has  an $U(1)_2$ charge sector and an $SU(2)_2/U(1)$ neutral sector that need to be consistently glued together\cite{moore-read-1991,fradkin-nayak-tsvelik-wilczek-1998,Dong2008,bishara-2008b,fradkin-nayak-2009}. 

Furthermore, the non-Abelian states have a richer excitation spectrum than their Abelian counterparts, which now contains:
\begin{itemize}
\item
Half-vortices (denoted by $\sigma$) with charge $q=e/4$ (fermionic case) and $q=e/2$ (bosonic case) and non-Abelian fractional (braid) statistics. 
\item
Vortices that are charge neutral Majorana fermions (denoted by $\psi$).
\item
Laughlin vortices with charge $e/m$ and abelian fractional statistics $\pi/m$
\end{itemize}
In summary, for the non-Abelian FQH states the effective theory has an Abelian charge sector, a $U(1)_m$ Chern-Simons gauge theory, and a neutral sector, which is described by a  non-Abelian $SU(2)_q$ Chern-Simons gauge theory with a $U(1)$ subgroup moded out (a ``coset''). The charge and neutral sectors are glued together by the requirement that the states thus obtained are local with respect to the electron (as the FQH fluid is an electron condensate)\cite{moore-read-1991}. 

Below we will use this construction to compute the entanglement entropy for Abelian and non-Abelian FQH states directly from the effective Chern-Simons gauge theory\cite{Dong2008}. Details of the construction of these effective field theories will be given in Ref.\cite{fradkin-nayak-2009}.

\subsection{Time Reversal Invariant Systems and Quantum Dimer Models}
\label{sec:time-reversal}

Quantum dimer models (QDM) are time reversal invariant lattice systems, originally proposed to describe the quantum frustration of antiferromagnetism in a doped Mott insulator\cite{Rokhsar1988}. In that framework, the dimer degrees of freedom are valence-bond singlet states of nearby spins on a lattice (for a detailed description see Ref.\cite{Fradkin1991}). Quantum dimer models have a special value of its couplings, known as the Rokhsar-Kivelson (RK) point, where the Hamiltonian can be shown  have an
exact ground state wave function has the short range ``resonating valence bond'' (RVB) form
\begin{equation}
\vert \Psi_{\rm RVB} \rangle=\sum_{ {\{} C {\}} } \vert C\rangle
,\qquad  {\{} C {\}} =\; \textrm{all dimer coverings of the lattice}
\end{equation}
where states represented by different dimer configurations are taken to be an orthonormal basis.\footnote{This condition is not satisfied for a systems with spin $1/2$ degrees of freedom but it is accurate in $SU(N)$ spin systems in the large $N$ limit\cite{read1991}.}
A special, and very useful, feature of this state is that the weight os a configuration, {\it i.e.\/} a particular covering of the lattice by dimers, is the same as the Gibbs weight of a {\em classical} dimer model on the same lattice. in particular, the {\em norm} of this wave function is {\em equal} (in this case) to the partition function of classical dimers on the same lattice. Using the `quantum-classical' connection it is straightforward to show that the equal-time correlation functions of the quantum theory can be computed from the correlation functions of the corresponding classical dimer model for suitably defined observables. Since the classical dimer model is integrable, the corresponding correlators are known. 

This approach has been generalized to systems with more complex degrees of freedom. One such generalization is  a quantum version of the Baxter (or eight vertex) model, which has ordered and topological phases separated by lines of fixed points of the Lifshitz universality class\cite{Ardonne2004}. Other generalizations\cite{Papanikolaou2007b,Castelnovo2005} include ``doped'' interacting dimer models. There are also generalizations with non-Abelian time-reversal invariant states based on quantum loop and net models at their respective `RK' point\cite{Freedman2004b,Levin2005,Fendley2005,Fendley2008b}.

We will not give a detailed description of the phase diagram of QDMs here. However, it will suffice to say that the RK point for a system on a {\em bipartite lattice} describes  quantum (multi) critical points (if not preempted by a first order transition), whose effective field theory has dynamical critical exponent $z=2$  and massless deconfined topological excitations (`spinons' and `visons')\cite{Rokhsar1988,Moessner2001,Moessner2002,Fradkin2004,vishwanath2004}.

On the other hand, on {\em non-bipartite lattices}, the QDM at the RK point is in a topological $\mathbb{Z}_2$ deconfined phase with massive spinons and visons and a topological $4$-fold ground state degeneracy on a torus\cite{Moessner2001,Moessner2002}. This state is closely related to Kitaev's Toric code\cite{Kitaev2003} and to the $\mathbb{Z}_2$ gauge theory at its (ultra) deconfined point\footnote{This is discussed in considerable detail in the Appendix of Ref.\cite{Papanikolaou2007}.}

In the next section we present the effective field theory of the QDMs at their quantum critical point: the quantum Lifshitz model.

\setcounter{footnote}{0}
\section{Scaling of Entanglement Entropy for the Quantum Lifshitz Universality Class}
\label{sec:entanglement-quantum-lifshitz}

\subsection[The Quantum Lifshitz Universality Class]{The Quantum Lifshitz Universality Class}
\label{sec:quantum-lifshitz}

The quantum Lifshitz model is the effective field theory of QDMs at criticality\cite{Henley1997,Moessner2002,Ardonne2004}. A quantum (as well as classical) dimer model on a bipartite lattice admits a {\em height representation} in terms of a set of integer-valued degrees of freedom (the `heights') residing on the sites of the dual lattice (see B. Nienhuis\cite{Nienhuis1987}). We will label the dual degrees of freedom by $h$. However, not all height configurations are allowed. Thus to a configuration on with a dimer on the $x$-link of the direct lattice it is assigned a height configuration on the dual lattice such that the heights grow by one unit in the link of the dual lattice does not cross the dimer on the direct lattice and drops by 4 unites if it does. In this fashion, the assigned configurations are in one-to-one correspondence to each other. Thus heights are defined mod 4. This rule will be violated if a a site does not have a dimer attached to it (it is `hole') and it corresponds to a topological excitation in the height picture.

 It is shown in Refs.\cite{Moessner2002,Ardonne2004} that the dual height model can be coarse-grained resulting in a model of a scalar field $\varphi$, compactified (to reflect the constraints of the height configurations) by the requirement that all allowed operators (including the Hamiltonian) be invariant under $\varphi \to \varphi +2 \pi r$, where $r$ is the compactification radius. The quantum hamiltonian for the resulting model is
 \begin{equation}
  H=\int d^2 x \biggl[ \frac{1}{2} \Pi^2 +\frac{1}{2} \left(\frac{k}{4\pi}\right)^2 \left(\nabla^2 \varphi \right)^2 \biggr]
\label{eq:QLM}
\end{equation}
where $\varphi$ and $\Pi$ obey canonical equal-time commutation relations, $\left[\varphi(\vec x),\Pi(\vec y)\right]=i \delta^2(\vec x - \vec y)$, and $k$ is a (so far) arbitrary parameter.

The term ``quantum Lifshitz model'' follows from the path-integral representation of this quantum theory in $2+1$ Euclidean (imaginary time) space-time whose {\em action} is
\begin{equation}
S=\int d^2x d\tau \biggr[\left(\partial_\tau \varphi\right)^2+\frac{1}{2} \left(\frac{k}{4\pi}\right)^2 \left(\nabla^2 \varphi \right)^2 \biggr]
\end{equation}
where $\tau$ is the imaginary time coordinate. This (Euclidean) action is the same as the free energy of a classical statistical mechanical system in 3D at a (classical) Lifshitz critical point between a uniform (but anisotropic) state and a modulated state\cite{chaikin1995}.

It is trivial to show that the ground state wave function $\Psi_0[\varphi]$ of the Hamiltonian of Eq.\eqref{eq:QLM}  is {\em scale invariant} and given by
\begin{equation}
\Psi_0 [\varphi] \propto
e^{\displaystyle{-\frac{k}{8\pi} \int d^2 x\;  \left(\nabla \varphi (\mathbf x)\right)^2}}
\label{eq:QLM-wf}
\end{equation}
which has the form of a local Gibbs weight in classical statistical mechanics. Indeed, it corresponds to the simplest critical classical system in 2D, the free boson of Gaussian model. Moreover, the norm of the 2D wave function is the partition function of this classical critical conformally invariant system!
\begin{equation}
\| \Psi_0 \|^2 = \int \mathcal{D} \varphi \; e^{\displaystyle{-\frac{k}{4\pi} \int d^2x\; (\nabla \varphi(\mathbf x))^2}} = ``Z''
\end{equation}
Furthermore, not only the norm (squared) of the wave function maps to the partition function of the classical critical system (wit the weight squared) but the observables themselves can be mapped. In particular the observables of the effective theory have the form of vertex operators with charges determined by the compactification condition. (For details and generalizations see Ref.\cite{Ardonne2004}.)

In summary, this construction is a mapping to a 2D Euclidean CFT. Under this mapping we find that
\begin{itemize}
\item
The amplitude of $\ket{\varphi}$ is the Gibbs weight of a Euclidean 2D free massless scalar field: scale invariant wave functions
\item
At these quantum critical points the ground state wave function is scale invariant
\item
The equal-time expectation value of operators in the quantum Lifshitz model are
given by correlators of the massless free boson conformal field
theory with central charge $c=1$.
\item
Time-dependent correlators: dynamical exponent  $z=2$.
\item
By matching the correlation functions  of the QDM at the RK point  and Lifshitz models, one finds that the parameter $k$ has to be chosen to be $k=r=1$. 
\item
For the 2D quantum Baxter wave function\cite{Ardonne2004}, one finds that $k$ varies continuously as a function of the Baxter weights.
\item
This construction generalizes to  states with non-Abelian braid statistics\cite{Fendley2005}.
\item
In general the resulting theory is a unitary Euclidean CFT.
\end{itemize}

\subsection{Entanglement Entropy and Classical Partition Functions}
\label{sec:partition}

\begin{figure}
\begin{center}
\includegraphics[width=0.4\textwidth]{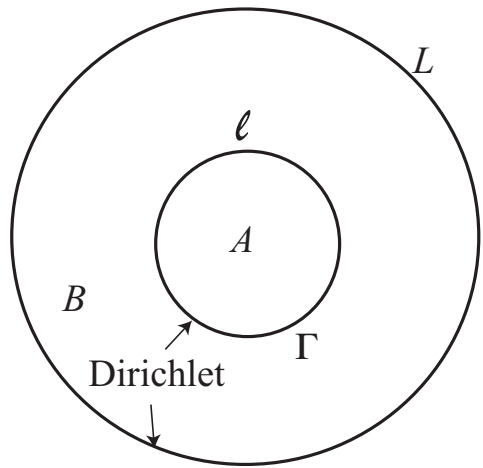}
\end{center}
\caption{The disk geometry}
\label{fig:diskAB}
\end{figure}

Let us compute the entanglement entropy for the ground state wave function $\Psi_0[\varphi]$ of the quantum Lifshitz model, Eq.\eqref{eq:QLM-wf}. We will follow in detail the arguments presented in Ref.\cite{Fradkin2006}. We will consider here the geometry of a disk (shown in \Fref{fig:diskAB})  with Dirichlet boundary conditions at infinity. Let region $A$ be a large disk of circumference $\ell$ , and $B$ be an annular region of inner circumference $\ell$ and outer circumference  $L$, such that $L \gg \ell \gg a$ (as before, here $a$ is the short distance cutoff). The common boundary $\Gamma$ between region $A$ and region $B$ is taken (for simplicity) to be a circle of circumference $\ell$.

For conformal quantum critical points, the Hilbert space has an orthonormal basis of states $|\{ \phi \} \rangle$ indexed by classical configurations $\{ \phi \}$, and the ground state $|\Psi_0\rangle$ of the bipartite system is determined by a CFT action $S$:
\beq
|\Psi_0\rangle = {1 \over \sqrt{Z_c}} \int\,\mathcal{D}\phi \,e^{-S(\{ \phi \}) /2} |\{\phi\}\rangle.
\eeq
Here 
\beq
Z_c=\int\,\mathcal{D}\phi\,e^{-S(\{\phi\})}
\eeq
and expectation values in this state reproduce CFT correlators.

To investigate the universal finite  terms in the entanglement entropy at 
2D conformal QCPs, we will rely on the approach described in the work of Fradkin and Moore\cite{Fradkin2006}.
Using the ``replica trick'' to compute the entanglement entropy\cite{Holzhey-1994,Calabrese2004}, they showed that the trace of the $n$th power of the reduced density matrix, $\textrm{tr} \rho_A^n$, where $\rho_A$ is
the (normalized) reduced density matrix of a
region $A$, with $A \subset B$ separated by the boundary $\Gamma$,
for the ground state $\Psi_0$ on $A \cup B$, 
is given by
\begin{equation}
\textrm{tr}\rho_A^n=  \frac{Z_n}{Z^n} =  \left(\frac{Z_A Z_B}{Z_{A \cup B}}\right)^{n-1}.
\label{eq:rhoA^n}
\end{equation} 
Here $Z_n$ is the partition function of $n$ copies of the equivalent 2D classical statistical mechanical 
system 
satisfying the constraint
that their degrees of freedom are identified on the boundary $\Gamma$, and
$Z^n$ is the partition function for $n$ decoupled systems. The partition functions on the  r.h.s of 
Eq.\eqref{eq:rhoA^n} are 
$Z_A=||\Psi_0^A||^2$ with support on region $A$ and $||\Psi_0^B||^2$ with support in region $B$, 
both satisfying 
generalized Dirichlet ({\it i.e.\/} fixed) boundary conditions on $\Gamma$ of $A$ and $B$, and 
$Z_{A \cup B}=||\Psi_0||^2$ is the norm squared for the full system.
The entanglement entropy $S$ is then obtained by an analytic continuation in $n$,  
\begin{equation}
S=-\textrm{tr}
\left(\rho_A \ln \rho_A\right) = - \lim_{n \to 1} \frac{\partial}{\partial n} 
\textrm{tr} \rho_A^n = - \log \left(\frac{Z_A Z_B}{Z_{A \cup B}}\right)
\label{eq:SFM}
\end{equation}
Hence, the computation of the entanglement entropy is reduced to the computation of a ratio of 
partition functions in a 2D classical
statistical mechanical problem, an Euclidean CFT in the case of a critical wave function, 
 each satisfying specific boundary conditions.

In order to construct  $\textrm{tr} \rho_A^n$,
  we need an expression for the matrix elements of the reduced density matrix 
  $ \me{\phi^A}{\rho_A}{{\phi^\prime}^A}$. 
Since the ground state wave function 
 is a local function of the field $\phi(x)$, 
a general matrix element of the reduced density matrix is a trace of the density
matrix of the pure state $\Psi_{GS}[\phi]$ over the degrees of freedom of the ``unobserved'' region $B$, 
denoted by $\phi^B(x)$. Hence the matrix elements of $\rho_A$ take the form
\begin{eqnarray}
&&\me{\phi^{A}}{ \hat{\rho}_{A}}{ {\phi^\prime}^A    }
= \nonumber \\
&&\frac{1}{Z} \int [D\phi^{B} ] \,\, e^{\displaystyle{-\left(\frac{1}{2} S^{A}(\phi^{A}) + 
\frac{1}{2} S^{A}({\phi^\prime}^A ) 
+S^B(\phi^B)\right)}},
\nonumber \\
&&
\end{eqnarray}
where the degrees of freedom satisfy the {\em boundary condition} at the common boundary
 $\Gamma$:
\begin{equation}
BC_\Gamma:\quad \phi^B|_\Gamma=\phi^A|_\Gamma={{\phi^\prime}^A}|_\Gamma.
\label{eq:BCphiGamma}
\end{equation}
Proceeding with the computation of 
$\textrm{tr}\rho_A^n$, it is immediate to see that the matrix product  requires the condition  $\phi^A_i={\phi^\prime}^A_{i-1}$
 for $i=1,\cdots,n$, and ${{\phi^\prime}^A_n}=\phi^A_1$ from the trace condition. 
Hence, $\textrm{tr}{\rho_A^n}$ 
takes the form
\begin{eqnarray}
\textrm{tr} \rho_A^n&\equiv& \frac{Z_n}{Z^n}
\nonumber \\
&=&\frac{1}{Z^n} \int_{BC_\Gamma} \prod_i D \phi_i^A D\phi_i^B \; e^{\displaystyle_{-\sum_{i=1}^n
\left(S(\phi_i^A)+S(\phi_i^B)\right)}}
\nonumber \\
&&
\label{eq:trrhoAn1}
\end{eqnarray}
{\em subject to the boundary condition $BC_\Gamma$} of Eq.\eqref{eq:BCphiGamma}. 
Notice that the numerator, $Z_n$ is the partition
function on $n$ systems whose degrees of freedom are identified in $\Gamma$ but are 
otherwise independent. Also notice the absence of the factors of $1/2$ in the exponentials 
of Eq.\eqref{eq:trrhoAn1}.

The other important
consideration is that the compactification condition requires that two fields that differ by 
$2\pi r$ be equivalent. Hence, the
boundary condition of Eq.\eqref{eq:BCphiGamma} is defined {\em modulo $2\pi r$}. 
(Equivalently, the proper form of the
degrees of freedom is $e^{i\phi}$.) This means that one can alternatively define 
$Z_n$ as a partition function for $n$ 
systems which are decoupled {\em in the bulk} but have a boundary coupling of the 
form (in the limit $\lambda_\Gamma \to \infty$,
which enforces the boundary condition)
\begin{equation}
S_\Gamma=-\oint_\Gamma \lambda_\Gamma \sum_{i=1}^n
\cos(\phi_i-\phi_{i+1}).
\label{eq:SGamma}
\end{equation}
 Here the fields $\phi_i$ extend over the entire region $A \cup B$. 
Thus, this problem maps onto a boundary CFT for a system with $n$ ``replicas'' 
coupled only through the boundary condition on the closed contour $\Gamma$, the boundary 
between the $A$ and $B$ regions.

For the special case of the free scalar field, one can simplify this further by taking 
linear combinations of the replica fields. 
Then the condition that the scalar fields $\phi_{i}$ agree with each other on $\Gamma$ 
can be satisfied by forming $n-1$ relative 
coordinates $\varphi_i\equiv  \phi_{i}-\phi_{i+1}$ ($i=1,\ldots,n-1$) 
that vanish ({\em mod} $2\pi r$) on $\Gamma$, and one ``center of mass 
coordinate'' field $\phi\equiv \frac{1}{\sqrt{n} } \sum_{i=1}^n \phi_{i} $ 
that is unaffected by the boundary $\Gamma$ (reflecting the fact that nothing physical 
takes place at $\Gamma$).
Hence, the computation of $\textrm{tr} \rho_A^n$ reduces  to the product of two partition functions: 
\begin{enumerate}
\item
 The partition function for the ``center of mass'' field $\phi$; since $\phi$ does not see the 
 boundary $\Gamma$, this is just the partition function $Z_{A \cup B}$ for a single field in the 
 entire system. 
 \item
  The partition function for the $n-1$ fields $\varphi_i$ which are independent from each other 
  and vanish ({\em mod} $2\pi r$ on $\Gamma$. We denote this by $\left({Z^D_\Gamma}\right)^{n-1}$. 
  However,  the fields $\varphi_i$ on the $A$ and $B$ regions are effectively decoupled from each other. 
  Hence, this partition function further factorizes to $Z^D_\Gamma=Z_A^D Z_B^D$, where $Z_A^D$ and $Z_B^D$ 
  are the partition functions for a single field $\phi$ on $A$ and $B$ respectively, 
  satisfying in each case Dirichlet (fixed) boundary conditions ({\em mod} $2\pi r$) 
  at their common boundary $\Gamma$.
  \end{enumerate}

For any CFT there exists a conformal boundary condition that generalizes the notion of the Dirichlet boundary condition in the free case (see below).  In terms of the partition functions $Z_D$, for a field in the whole system $A \cup B$ that vanishes at the boundary, and $Z_F$, for a field that is free at the boundary,
\beq
{\rm Tr}\,\rho_A^n = {Z_D^{n-1} Z_F \over {Z_F}^n} = \left({Z_D \over Z_F}\right)^{n-1}
\label{eq:SFM}
\eeq
and therefore
\beq
S = -\log {Z_D \over Z_F} = - \log {Z_D^A Z_D^B \over Z_F}.
\eeq
In the last equality,  the Dirichlet boundary condition at the boundary was used to split the partition function into contributions from $A$ and $B$, each including the boundary with Dirichlet boundary conditions.  Finally, the entanglement entropy for a general conformal quantum critical point is just the dimensionless free energy difference induced by the partition in the associated CFT\cite{Fradkin2006}
\beq
S = F_A + F_B - F_{A \cup B}.
\label{composition}
\eeq
where $F_A$, $F_B$ and $F_{A \cup B}$ are the {\em free energies} of the CFT associated with the ground state wave function, with specific boundary conditions. For the case of the quantum Lifshitz wave function, \Eref{eq:QLM-wf}, $F_A$ obeys Dirichlet boundary conditions on $\Gamma$, $F_B$ obeys Dirichlet boundary conditions on the inner boundary $\Gamma$ and as-yet-unspecified boundary conditions on its outer boundary, and $F_{A \cup B}$ obeys the same boundary conditions on its (only outer) boundary as $F_B$. 

We have thus succeeded in expressing the entanglement entropy in terms of a combination of free energies of 2D Euclidean CFTs satisfying specified boundary conditions. \Eref{composition} is actually of general validity. For the case of a general CFT, Dirichlet boundary conditions are replaced by fixed boundary conditions, which generally implies that the boundary state is in the conformal block of the identity\cite{Cardy1986a,Cardy1989,hsu-2009}. Hence, the problem of computing the entanglement entropies of conformal wave functions reduces to a problem in boundary Euclidean CFT in 2D.

To determine the scaling of the entanglement entropy with the linear size of the region being observed we thus need to know the same scaling for the free energies of \Eref{composition}. The latter is a problem that has been studied for quite a long time and much is known about it.  For a large bounded region of linear size L and smooth boundary, $F$ obeys the `Mark Kac Law'\cite{Kac1966}\footnote{`Can you hear the shape of a drum?'}
\beq
F=\alpha L^2 + \beta L -\frac{c}{6} \chi \ln L + O(1), 
\eeq
a general result due to Cardy and Peschel\cite{Cardy1988}.
Here, $\alpha$ and $\beta$ are non-universal constants, $c$ is the central charge of the CFT, and $\chi$ is the Euler characteristic of the region in consideration:
\beq
\chi=2 - 2h -b,
\label{eq:euler}
\eeq
where $h$ and $b$ are the number of handles and the number of boundaries of the region. 

Hence, the logarithmically dependent term in the entanglement entropy is
\beq
\Delta S=-\frac{c}{6} \left(\chi_A+\chi_B-\chi_{A \cup B} \right) \log L
\eeq
We will see below that the  $O(1)$ correction has a {\em universal piece} related to the ``boundary entropy'' of Affleck and Ludwig in boundary CFT\cite{Affleck1991}.

For regions $A \subseteq B$ the coefficient of the $\log L$ term {vanishes} since in this case there is no net change in the Euler characteristic:
\beq
\chi_A+\chi_B=\chi_{A \cup B} \Rightarrow \Delta S=0
\eeq
On the other hand if under some {\em physical process} $A$ and $B$ become physically separate and have no common intersection, it follows that $\chi_A+\chi_B-\chi_{A \cup B}\neq0$. Hence, if the system physically splits in
two disjoint parts, then  there is a  $\log L$ term in the entanglement entropy.
Logarithmic terms in the entanglement entropy also arise if neither $A$ nor $B$ is a subset of the other region, yet  share a common boundary. In this case, there is a 
$\log L$ term whose coefficient is determined by the angles at the intersections. Finally, if the boundary of $A$ is not smooth, then the coefficient depends on the angles $\alpha_i$ for both regions where the boundary $\Gamma$ is singular. (For details, see Ref\cite{Fradkin2006}).

\subsection{Universal Finite Contributions to the Entanglement Entropy: The Quantum Lifshitz Case}
\label{sec:finite}

In this section we will consider the interesting (and generic) case in which the logarithmic terms are cancel out, by the argument given in the previous section. This problem  was discussed in detail by Hsu et al\cite{hsu-2009} and we follow their treatment closely.

If the coefficient of the logarithmic term vanishes, the $O(1)$ becomes universal.
The finite term is determined by the contributions of the winding modes to the respective partition functions. These partition functions have been computed and studied extensively in the CFT literature\cite{Ginsparg1988,yellow}.
The result depends on the topology of the surface and on the properties of the CFT associated with the wave function.

For the Quantum Lifshitz universality class on a {\em cylinder}, with $L_{A,B} \gg \ell$, the partition function for  a boson with  compactification radius $R$ on cylinder of length $L$ and circumference $\ell$ with Dirichlet boundary conditions on both ends, which is well known:\cite{Fendley1994} 
\begin{equation}
Z_{DD}(L,\ell)=\mathcal{N}\; \frac{1}{R} \frac{\vartheta_3\left(\frac{2\tau}{R^2}\right)}{\eta(q^2)}
\label{eq:ZDD-cylinder}
\end{equation}
where $R=\sqrt{2r^2k}$ is the effective compactification radius (as before), and $\mathcal{N}$ is a non-universal regularization-dependent prefactor, responsible for the area  and 
perimeter dependent terms in the free energy. 
(There are no logarithmic terms for a cylinder or a torus as their Euler characteristic $\chi$ vanishes.)
In Eq.\eqref{eq:ZDD-cylinder} $\tau=i\frac{L}{\ell}$ is the modular parameter, encoding the geometry of the cylinder, 
and $q=e^{2\pi i \tau}$. The elliptic theta-function $\vartheta_3(\tau)$ and the Dedekind eta-function $\eta(q)$
are given by
\begin{equation}
\vartheta_3(\tau)=\sum_{n=-\infty}^\infty q^{\frac{n^2}{2}}, \quad
\eta(q)=q^{\frac{1}{24}} \prod_{n=1}^\infty (1-q^n).
\label{eq:theta3-eta}
\end{equation}
The important feature of Eq.\eqref{eq:ZDD-cylinder} is  the factor $1/R$, the contribution of the winding modes of the compactified boson on the cylinder 
with Dirichlet boundary conditions. 

Putting it all together, it is straightforward  to find an expression for the entanglement entropy using Eq.\eqref{eq:SFM}. 
In general, the entanglement entropy depends on the geometry ({\it e.g.\/} the aspect ratios $L/\ell$) of the cylinders, 
encoded in ratios of theta and eta functions. However, 
in the limit $L_A\gg \ell$, in which  the length of the cylinders are long compared to their circumference, 
the entanglement entropy given by Eq.\eqref{eq:S-cyl} and Eq.\eqref{eq:ZDD-cylinder} takes a simple form
\begin{equation}
S=\mu \ell +  \ln R,
\label{eq:entropy-cylinder}
\end{equation}
where $\mu$ is a non-universal constant that depending on the regularization-dependent pre-factor $\mathcal{N}$
of Eq.\eqref{eq:ZDD-cylinder}.  Hence, there is a $\mathcal{O}(1)$ universal contribution to the entanglement entropy 
 $\gamma_{QCP}= \ln R$ for the cylindrical geometry.  
The explicit dependence of $\gamma_{QCP}$ on the effective effective compactification radius $R=\sqrt{2kr^2}$ shows that it is determined by the winding modes of the compactified boson 
and thus it  is a universal quantity determined by the topology of the surface.
In particular we find that the universal piece of the entanglement entropy, $\gamma_{QCP}$, for a compactified boson is a 
 continuous function of the radius $R$, a consequence of the existence of an exactly marginal operator at this QCP. 
 We find the similar relations for all topologies we considered.

Hence, for the cylinder geometry the $O(1)$ universal term $S_{\rm cylinder}$ equals
\beq
S_{\rm cylinder}=\log R
\eeq
 where $R=\sqrt{2 k r^2}$ is the compactification radius of the associated CFT.\footnote{Notice that the compactification radius of the associated Euclidean CFT is not the same as that of the $2+1$-dimensional theory.}
\begin{figure}[h!]
\begin{center}
\includegraphics[width=0.5\textwidth]{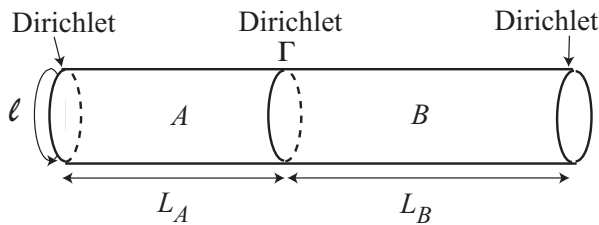}
\end{center}
\caption{The cylinder geometry.}
\label{fig:cylinderAB}
\end{figure}
For the RK quantum dimer model, $k=r=1$ and we obtain $R=\sqrt{2}$. 

Hence, for the Quantum Lifshitz model on a cylinder, the universal finite term in the entanglement entropy is $ \log \sqrt{2}$. It differs from its value $- \ln 2$  in the nearby topological phase, which as a topological phase  is in the universality class of Kitaev's toric code.

\begin{figure}
\begin{center}
\includegraphics[width=0.5\textwidth]{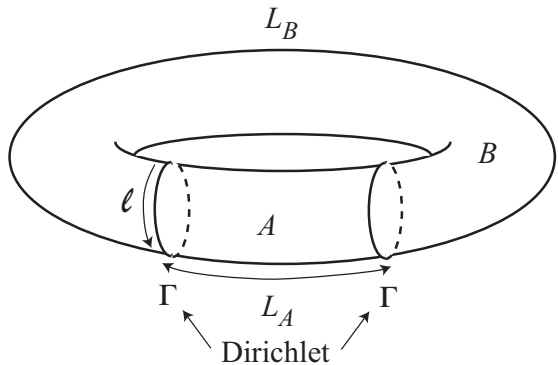}
\end{center}
\caption{The torus geometry.}
\label{fig:torusAB}
\end{figure}

The case of  the case of quantum Lifshitz model on a torus can be analyzed similarly. We now consider the case in which the full system $A\cup B$ is a torus for which the real part of the modulus $L/\ell \gg 1$, 
as shown in Fig.\ref{fig:torusAB}. The two subsystems, $A$ and $B$ are now two cylinders, 
of length $L_A$ and $L_B$ respectively ($L=L_A+L_B$), 
both with the same circumference $\ell$. We will thus need the partition function 
on a torus and on two cylinders (with both ends of the cylinders obeying Dirichlet boundary conditions.) 
The trace $\textrm{tr} \rho_A^n$ now becomes
\begin{equation}
	\textrm{tr } \rho_{A}^{n} = 
	\left( \frac{Z^{A}_{DD}(L_A,\ell) Z^{B}_{DD}\left(L_B,\ell\right)}{Z^{A\cup B}_{\rm torus}(L,\ell) } \right)^{n-1}. 
\end{equation}
The partition functions for the two cylinders, $A$ and $B$ has the form of Eq. \eqref{eq:ZDD-cylinder}. 
The partition function
for the torus is\cite{yellow,Ginsparg1988} 
\begin{equation}
Z_{\rm torus}(L,\ell) = \left(Z_{\rm cylinder}^{NN}\left(\frac{L}{2},\ell \right)\right)^2,
\label{eq:torus-cylinder}
\end{equation}
where $Z_{\rm cylinder}^{NN}(\frac{L}{2},\ell)$ is the partition function on a cylinder of length $\frac{L}{2}$ and
circumference $\ell$, with Neumann boundary conditions at both ends:
\begin{equation}
Z_{\rm cylinder}^{NN}\left(\frac{L}{2},\ell \right)=\mathcal{N}\; \sqrt{\frac{kr^2}{2}}\; \frac{\vartheta_3\left(\tau k r^2\right)}{\eta(q^2)},
\label{eq:}
\end{equation}
where $\tau=i\frac{L}{\ell}$ and $q=\exp(2\pi i \tau)$. 

In the limit $L_A \gg \ell \gg a$ and $L_B \gg \ell \gg a$, the entanglement entropy for the toroidal geometry is
\begin{equation}
S=\mu \ell +
2 \ln \left(\frac{R^2}{2}\right).
\label{eq:entropy-torus}
\end{equation}
Hence, for the toroidal geometry, the universal term is $\gamma_{QCP}=2 \ln \left(kr^2 \right)=2 \ln (R^2/2)$. 
In Eq.\eqref{eq:entropy-torus} $\mu$ is, once again, a non-universal factor which depends on both the short distance regularization and boundary
conditions (in fact, it is not equal to the constant we also called ``$\mu$'' in the entanglement entropy 
for the case of the
cylinder, Eq.\eqref{eq:entropy-cylinder}.)  As was the case for the cylindrical geometry, 
in the case of the torus
$\gamma_{QCP}$ is also determined by the contribution of the zero modes of the compactified boson to the partition functions.
Thus, here too, $\gamma_{QCP}$ depends on the 
effective boson radius $R=\sqrt{2kr^2}$. However, the different values of 
$\gamma_{QCP}$ in Eq.\eqref{eq:entropy-torus} and Eq.\eqref{eq:entropy-cylinder}
is due to the fact that on the torus all
three partition functions have contributions from the zero modes.

For a disk, with the geometry of \Fref{fig:diskAB}, the finite universal term $S_{\rm disk}$ depends also on the {\em aspect ratio}, and the universal finite term is\cite{hsu-2009}
\beq
S_{\rm disk}=\frac{1}{2} \ln \left[\frac{1}{\pi} \ln \left(\frac{L}{\ell}\right)\right]+\ln R
\eeq

We close this section with the application of these results to several problems of interest. As noted in previous sections several systems of interest can be mapped to the quantum Lifshitz model for an appropriate choice of the parameter $k$ and the compactification radius $r$. Systems of this type include: a) the quantum dimer model\cite{Rokhsar1988}, for which $k=r=1$, b) a generalized `interacting' quantum dimer model\cite{Castelnovo2005,Papanikolaou2007b}, for which $k$ varies continuously but $r$ is fixed to 1, and c) the quantum eight vertex model of Ref.\cite{Ardonne2004} for which $r=1$ and $k$ varies continuously (long the six vertex line) as (here $c$ is a Baxter weight)
\begin{equation}
\frac{\pi}{2k}=\cot^{-1}\sqrt{\frac{4}{c^4}-1}
\label{eq:6v}
\end{equation}
In all three cases we have mapped the problem to an Euclidean  boson CFT in 2D with compactification radius  $R=\sqrt{2kr^2}$. In all three systems, there is a perturbation that destabilizes the fixed line and drives the system immediately  into a topological phase. For the dimer models, the relevant perturbation that drives the system into the topological phase are dimers on both sublattices\cite{Ardonne2004}, whereas in the case of the quantum Baxter model the Baxter weights that break the continuous symmetry of the six-vertex model (down to a $\mathbb{Z}_2$) are the responsible for this transition. As shown in Ref.\cite{Papanikolaou2007}, the universal term of the entanglement entropy is the same (and hence {\em universal})  in the entire topological phase, taking the value $-\ln 2$, associated with its value at the Kitaev point, the stable fixed point of the topological phase. Hence, while in the topological phase the universal term has a fixed (and negative) value, along the critical lines the universal term of the entanglement entropy varies continuously as a function of the parameter $k$.\footnote{This should not be a surprise since the critical lines exist due to the presence of a exactly {\em marginal} operator in the theory.} Hence, in general, we should expect that the universal term of the entanglement entropy may jump at phase transitions.\footnote{Indeed, in an ordered phase, we expect the finite universal terms to vanish.}

\subsection{Universal Finite Contributions to the Entanglement Entropy: General Case}
\label{sec:finite-general}

The results of the previous sections generalize for a general conformally invariant wave function defined by the Gibbs weights of a 2D (Euclidean) rational conformal field theory (RCFT), {\em e.g.} the Ising model, quantum loop and net models. Once again, the norm (squared) of the wave function becomes a partition function of an associated 2D Euclidean CFT. However since these are no longer free field theories the identification of the CFT is a little more involved\cite{Fendley2005}. 

In this section we will compute the entanglement entropy for wave functions whose weights are associated with a RCFT, following the work of Ref.\cite{hsu-2009}. Here we rely on very basic and standard properties of CFTs which can be found in many texts, see {\it e.g.\/} Ref.\cite{yellow}.
A 2D RCFT has a set of {\em primary fields} $\Phi_a$   have an operator product expansion (OPE) of the form
\beq
\Phi_a \times \Phi_b=\sum_j N^j_{ab} \Phi_j
\eeq
 where the integer-valued coefficients $N^j_{ab}$ are known as the fusion coefficients. On the other hand, each primary field define a set of conformally invariant boundary conditions labelled by $a$. This defines a boundary RCFT\cite{Cardy1988}.
 
A rational CFT (RCFT) is a CFT with a finite number of primaries. Under a modular transformation the partition function of an RCFT with  boundary conditions specified by the action of its primary fields, transforms linearly.  These modular transformation laws are encoded in the modular $\mathcal{S}$-matrix which is related to the fusion coefficients by the Verlinde formula\cite{Verlinde1988}
\beq
N^j_{ab}=\sum_i \frac{S^i_j S^i_a S^b_i}{S^i_0}
\eeq
 The partition function for a RCFT on a cylinder 
of length $L$ and circumference $\ell$, with boundary conditions $a$ and $b$ on the left and right ends respectively, 
$Z_{a/b}$, can be expressed in terms of the characters $\chi_i$ of the RCFT:
\begin{equation}
Z_{a/b}=\sum_j N^j_{ab} \chi_j\left(e^{\displaystyle{-\pi \ell/L}}\right),
\label{eq:Zbc}
\end{equation}
where $N^j_{ab}$ are the fusion coefficients.

The Virasoro characters $\chi_j$ are given by the trace over the descendants  $\ket{\Phi_j}$ of the highest weight state, which are obtained by acting on it 
with the Virasoro generators $\hat{L}_{-n}$ ($n>0$):
\begin{equation}
\chi_j(e^{-\pi \ell/L})=e^{{\pi \ell c }/{24 L}} \; 
\textrm{tr}_a\left(e^{-\frac{\pi \ell}{L} \hat{L}_0}\right),
\label{eq:characters}
\end{equation}
where $c$ is the central charge of the CFT, $\hat{L}_0$ is the $n=0$ Virasoro generator. Here the modular parameter is $\tau\equiv i\ell/2L$.
Under a modular transformation $\tau\rightarrow-1/\tau$, which exchanges the Euclidean ``space'' and ``time''  
dimensions of the cylinder ({\it i.e.\/} it flips the cylinder from the ``horizontal'' to the ``vertical'' position), 
the characters transform as
\begin{equation}
\chi_i\left(e^{\displaystyle{-\pi \ell/L}}\right)=S^j_i\; \chi_j\left(e^{\displaystyle{-4\pi L/\ell}}\right),
\label{eq:modular}
\end{equation}
where $S^j_i$ is the {\em modular $\mathcal{S}$-matrix} of the RCFT.  The modular $\mathcal{S}$-matrix and
the fusion coefficients are related by the Verlinde formula.

The limit of interest here is, once again, $L \gg \ell$. Under a modular transformation, the partition function 
of Eq.\eqref{eq:Zbc} becomes
\begin{equation}
Z_{a/b}=\sum_{i,j} N^i_{ab} \; S^j_i\; \chi_j\left(e^{\displaystyle{-4\pi L/\ell}}\right).
\label{eq:Zbc2}
\end{equation}
In the limit $\frac{\ell}{L} \to 0$, $Z_{a/b}$ is dominated by the 
 the descendants of the identity $\bf{1}$ (up to exponentially small corrections). Hence, in this limit,
\begin{equation}
Z_{a/b} \to \sum_i N^i_{ab} \; S_i^0 \; \chi_0\left(e^{-4\pi L/\ell}\right) \to 
e^{{\frac{\pi L c}{6\ell}}} \; \sum_i N^i_{ab} \; S^0_i
\label{eq:lowT}
\end{equation}
and  $\ln Z_{a/b}$ becomes
\begin{equation}
\ln Z_{a/b}=\frac{\pi L c}{6\ell}+\ln g_{ab},
\label{eq:Zbc-g}
\end{equation} 
dropping UV singular (non-universal) terms. The quantity $\ln g_{ab}$ in Eq.\eqref{eq:Zbc-g} is the 
{\em boundary entropy} of a boundary RCFT introduced by Affleck and Ludwig\cite{Affleck1991}, where the ``ground state degeneracy'' 
$g_{ab}$ is given by
\begin{equation}
g_{ab}=\sum_i N^i_{ab} S_i^0.
\label{eq:g}
\end{equation}
  
For a cylindrical geometry, the entanglement entropy is
\beq
S=-\log\left(\frac{Z_D^A Z_D^B}{Z_{A\cup B}}\right)=\mu \ell -\ln \left(\frac{g_{a0}g_{0b}}{g_{ab}}
\right)
\eeq
 This is the main result , which shows that $\gamma_{QCP}$ is in general determined by the OPE coefficients $N_{ba}^c$
(which encode the boundary conditions on the partition functions) and by the modular $\mathcal{S}$-matrix, $S_i^j$, of the RCFT associated with
the {\em norm squared of the many-body wave function} at the given QCP. 

It is important to note that it is also possible to define a unitary $\mathcal{S}$-matrix that governs the transformation properties
of the {\em wave function} itself under a modular transformation. This modular $\mathcal{S}$-matrix plays a central role in 2D topological phases and in topological field theories.\cite{Witten1989,Kitaev2006a,Bonderson2006b} However, only for topological theories these two $S$-matrices are the same and in general are different or even not be defined at all!)  We will come back to this issue in the discussion section.

A particularly simple result is obtained for the case of a cylinder with fixed boundary conditions on both ends. In this case, $Z_A$,
$Z_B$ and $Z_{A \cup B}$ are cylinders with fixed boundary conditions, and hence the boundary states for all three cases are in the
conformal block of the identity ${\bf 1}$.  Since in this case the only non-vanishing OPE coefficient is $N_{0 0}^0=1$, the
universal term of the entanglement entropy, $\gamma_{QCP}$, depends only on the element $S_0^0$ of the modular $\mathcal{S}$-matrix of the RCFT:
\begin{equation}
\gamma_{QCP}=-\ln S_0^0.
\label{eq:gammaQCP-simple}
\end{equation}

Hsu et al\cite{hsu-2009} give a detail application of these results to the computation of the entanglement entropy in a few interesting non-trivial systems: a)  a 2D quantum system whose wave function has the weights of the Gibbs weights of the 2D Ising model, b) quantum loop models, and 3) quantum net models (including the chromatic polynomial model ). For details see Ref.\cite{hsu-2009}.

\section{Quantum Entanglement Entropy and Chern-Simons Gauge Theory}
\label{sec:chern-simons}

We now turn to the problem of computing the entanglement entropy in topological phases using the effective topological field theory. Here we will focus on the case of Chern-Simons gauge theories and the related FQH fluids. We will follow closely the results of Dong et al\cite{Dong2008}.

In previous sections we showed that the FQH wave functions represent topological fluids with a finite correlation length 
$\xi \propto \ell$ ($\ell$ is the magnetic length). 
Recently, the entanglement entropy of FQH states has be computed numerically  by K. Schoutens et al\cite{Haque2007,Zozulya2007,haque-2009} and by Li and Haldane\cite{Haldane2008}. 
Here we will show\cite{Dong2008} that one can compute the entanglement entropy directly from the effective field theory of all FQH states:
Chern-Simons gauge theory. This result can be applied directly to all known FQH states. Since it uses the effective topological field theory, it computes only the topological invariant piece of the entanglement entropy. The resulting universal topological entanglement entropy is given in terms of the modular $\mathcal{S}$-matrix of the effective Chern-Simons theory and of its conformal blocks.

In Ref.\cite{Dong2008} we computed the entanglement entropy for a level $k$ Chern-Simons theory on a smooth manifold with any number of handles, using  the seminal results of Witten\cite{Witten1989}
for the Chern-Simons theory. The action of a (non-Abelian) Chern-Simons gauge theory is
\beq
S(A)=\frac{k}{4\pi} \int \textrm{Tr} \left(A \wedge dA+\frac{2}{3} A\wedge A \wedge A \right)
\label{eq:action-cs}
\eeq
where, as usual, $A_\mu$ is a vector field taking values in the algebra of a (compact) gauge group $G$. Here we will be primarily interested in the vase of $G=SU(2)$.

We will need a few important results on the structure of the Chern-Simons theory and its solution. 
\begin{itemize}
\item
Following Witten\cite{Witten1989}, we realize the states on a closed 2D surface as a path integral over a 3D volume.
\item
 Witten showed that the Chern-Simons states on a spatial manifold $\Sigma$ (which we will take to be closed) are in one-to-one correspondence with the conformal blocks of a Wess-Zumino-Witten (WZW) CFT.
 \item
  He also showed that
the ground state degeneracy depends on the level $k$ and on the topology of the surface $\Sigma$.
\item
The partition functions ({\it i.e.\/} the value of the path integral) depend on the matrix elements of the modular $\mathcal{S}$-matrix, {\it e.g.\/} the partition function on $S^3$ with a Wilson loop in representation $\rho_j$ is
\beq
Z(S^3,\rho_j)=\mathcal{S}_0^j
\eeq
\item
Here the modular $\mathcal{S}$-matrix defines how the  degenerate ground states on a torus transform under a modular transformation
\end{itemize}

We will need some properties of the modular $S$ matrix and of the conformal blocks. For the gauge group $U(1)_m$, $n=0, \ldots, m-1$, the modular $S$ matrix is
\beq
{{\cal S}_{[n']}}^{[n]} = \frac{1}{\sqrt{m}}e^{2\pi inn'/m}
\eeq
 whereas for the gauge group $SU(2)_k$, $j,j'=0,1/2,\ldots,k/2$, the modular $S$ matrix is
 \beq
 {S^{(k)}_j}^{j'}=\sqrt{\frac{2}{k+2}} \; \sin \left(\pi \frac{(2j+1)(2j'+1)}{k+2}\right)
 \eeq
 We also need the definition of the {\em quantum dimensions} $d_j$
 \beq
 d_j=\frac{\mathcal{S}_0^j}{\mathcal{S}_{00}}, \quad \mathcal{D} \equiv \sqrt{\sum_j |d_j|^2}=\frac{1}{\mathcal{S}_{00}}
 \eeq
 which measure the rate of growth of the degenerate Hilbert 
 spaces of particles labeled by the representation $\rho_j$.

The Chern-Simons path integral (its partition function) on various manifolds can be reduced to its computation on a sphere $S^3$ using the method of (Chern-Simons) surgeries\cite{Witten1989}. Using surgeries it is shown that
 if a 3-manifold $M$ is the connected sum of two 3-manifolds $M_1$ and $M_2$ joined along an $S^2$, then the Chern-Simons partition functions on these manifolds are related by
\beq
Z(M) Z(S^3)=Z(M_1) Z(M_2)
\eeq
 In particular, if $M$ is $M_1$ and $M_2$ joined along $n$ $S^2$'s, the resulting partition function is
\beq
Z(M)=\frac{Z(M_1) Z(M_2)}{Z(S^3)^n}
\eeq

\begin{figure}[h!]
	\begin{center}
\includegraphics[width=0.48\textwidth]{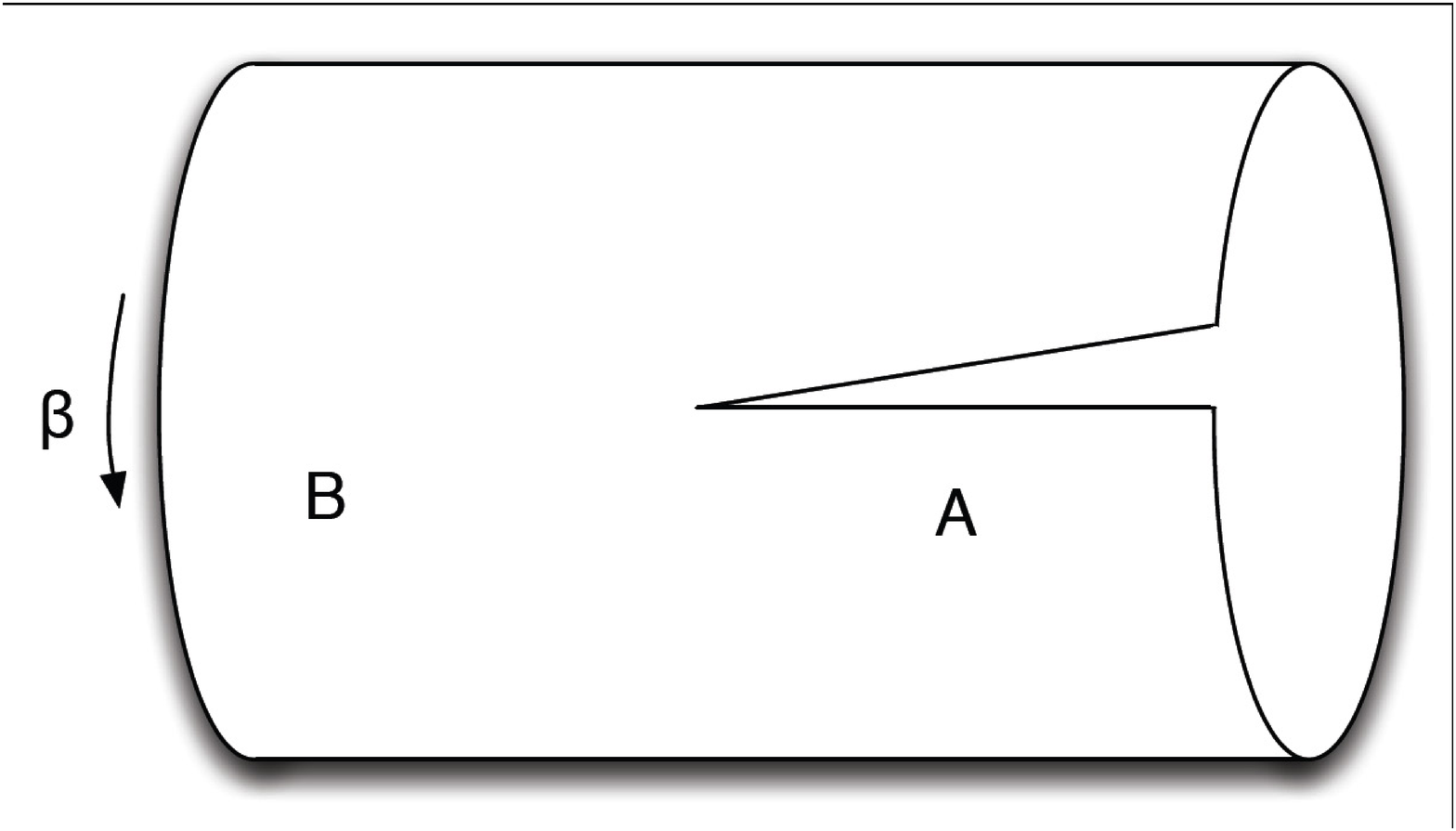}
\includegraphics[width=0.48\textwidth]{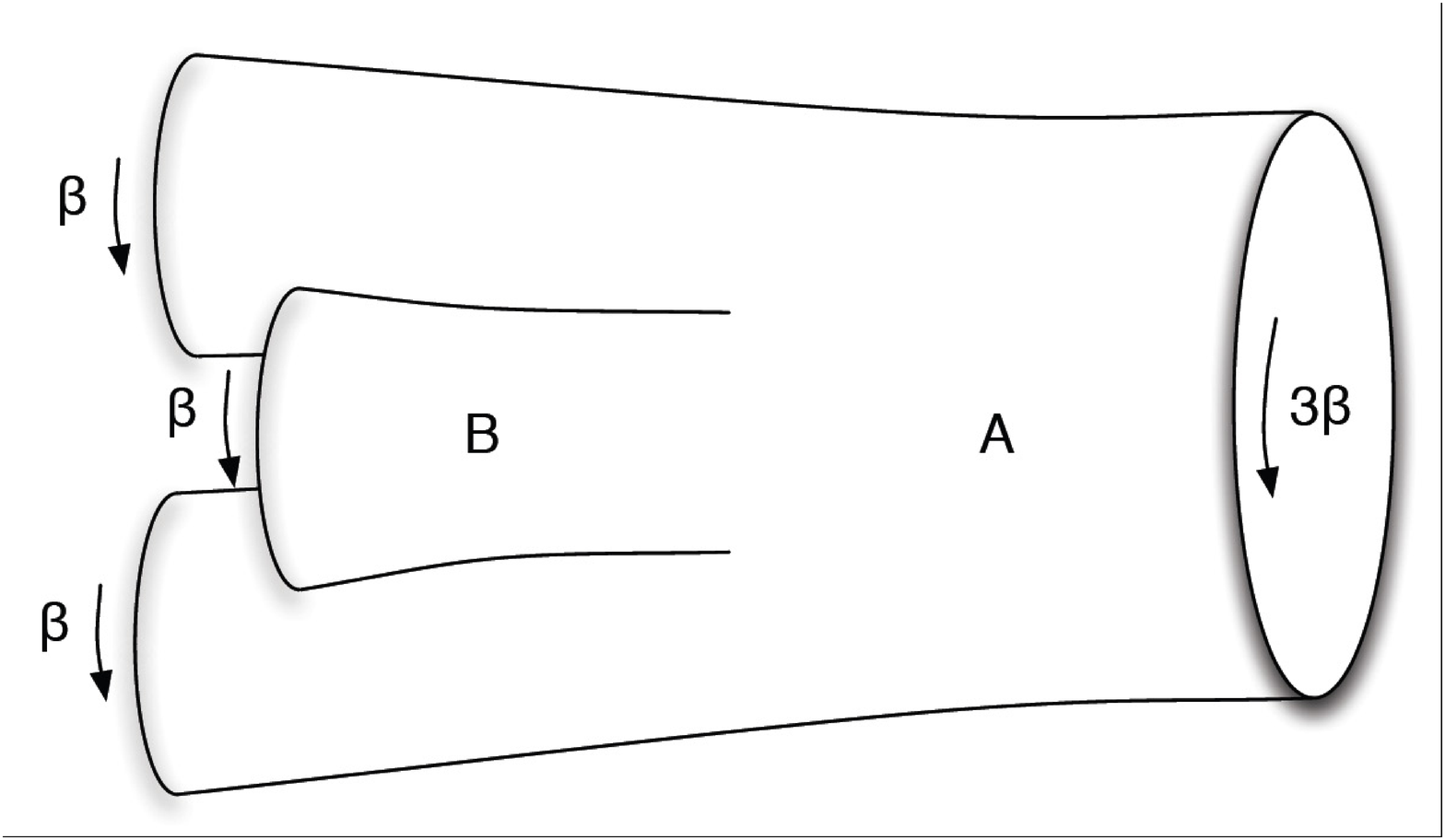}
\end{center}
\caption{Left: the cut-manifold defining the reduced density matrix $\rho_A$. Right: the foliated manifold used to compute $\textrm{Tr} \rho_A^3$. Here $\beta \to \infty$ is the inverse temperature.}
\label{fig:trrho3}
\end{figure}

We will compute the entanglement entropy for Chern-Simons theory using the path integral approach of Calabrese and Cardy\cite{Calabrese2004}, suitably adapted for the system at hand by Dong et al\cite{Dong2008}. In other terms, one uses a path-integral to compute the $n$th power of the reduced density matrix. This leads to a ``foliated'' 3D manifold shown  in \Fref{fig:trrho3} (for $n=3$)

Let us consider first the simplest case, in which the spatial manifold is a sphere, $\Sigma=S^2$, and hence the space-time manifold is just a 3-sphere, $\Sigma \times S^1 \cong S^3$. The Hilbert space on $S^3$ is one-dimensional.  
Using the method of surgeries Dong et al\cite{Dong2008} considered the case of  $S^2$ with one $A-B$ boundary ({\it i.e.\/} the observed region is a hemisphere). The two regions $A$ and $B$ are two hemispheres (disks). The 3-geometry is a ball.

 To construct $\textrm{tr} \hat \rho_A^n$ we glue $2n$ such pieces together. When glued together $S^2$ rotated about the axis which has the topology of $S^3$. For $n>2$, the $S^2$ is obtained by sequentially gluing $2n$ disks, we find that the (normalized) trace of $\rho_A^n$ is
\beq
\frac{\textrm{tr} \rho^n_{A(S^2,1)}}{\left(\textrm{tr} \rho_{A(S^2,1)}\right)^n}=\frac{Z(S^3)}{\left(Z(S^3)\right)^n}=\left(Z(S^3)\right)^{1-n}=\mathcal{S}_{00}^{1-n}
\eeq
In the replica limit, $n \to 1$, we obtain for the entanglement entropy
\beq
 S_A^{(S^2,1)}=\ln \mathcal{S}_{00}=-\ln \mathcal{D}
\eeq
which is the well known result of Kitaev and Preskill\cite{Kitaev2006a}, and Levin and Wen\cite{Levin2006} for the universal topological entanglement entropy.
This result  also holds for surfaces with arbitrary topology provided the region being observed $A$ is topologically trivial, regardless of the pure state labeled by the representations $\rho_j$.
For the case of a sphere $S^2$ and a disconnected connected region $A$ with $M$ boundaries we find
$ S_A^{(S^2,M)}=M\ln \mathcal{S}_{00}=-M \ln \mathcal{D}$.

\begin{figure}[h!]
\begin{center}
\includegraphics[width=0.9\textwidth]{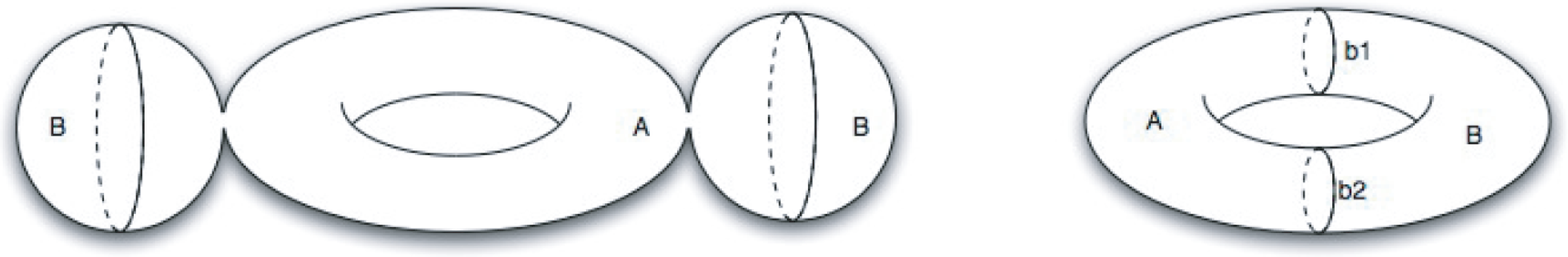}
\end{center}
\caption{Two cases for the torus.}
\label{fig:torus-cs}
\end{figure}

Let us compute the entanglement entropy for a Chern-Simons theory on the torus $T^2$ with more than one $A-B$ boundary. For a torus $T^2$ split into two regions with more than one (say two) boundary, we have two cases, shown in \Fref{fig:torus-cs}.

For the trivial state (no Wilson loop) the entropy is the same in both cases, 
\beq
S_A(T^2,2)=2 \ln \mathcal{S}_{00}
\eeq
If there is a Wilson loop with a non-trivial representation $\rho_j$, we obtain the same result for the case depicted in \Fref{fig:torus-cs} (left). But, for the case of \Fref{fig:torus-cs} (right), for a Wilson loop in representation $\rho$  we obtain instead,
\beq
S_A(T^2,2,\rho)=2\ln S_{0\rho}
\eeq
For a state which is a linear superposition, $\ket{\psi}=\sum_\rho \psi_\rho \ket{\rho}$, we find 
\beq
S_A(T^2,2,\psi)=2 \ln \mathcal{S}_{00}-\sum_\rho d_\rho^2 \left(\frac{|\psi_\rho|^2}{d_\rho^2} \ln \frac{|\psi_\rho|^2}{d_\rho^2}\right)
\eeq
Clearly, the entanglement entropy now depends not only on the effective quantum dimension $\mathcal{D}=\mathcal{S}_{00}^{-1}$ but also on the quantum dimension of the excitation labeled by the representation $\rho$, as well as on which particular linear combination of the degenerate vacua on the torus in which the system is prepared.

Let us consider now the computation of the entanglement entropy in the presence of quasiparticles, {\it i.e.\/} for a manifold with punctures each carrying a specific representation label.
Let us consider the case of four quasiparticles on $S^2$: $S^2$ with four punctures.
We will consider first the case of the Chern-Simons gauge theory $SU(N)_k$, with $N \geq 2$ and $k \geq 2$, with two punctures carrying fundamental $\hat \alpha$ and 2 anti-fundamental $\hat \alpha^*$ representations.

\begin{figure}[h!]
\begin{center}
\includegraphics[width=0.75\textwidth]{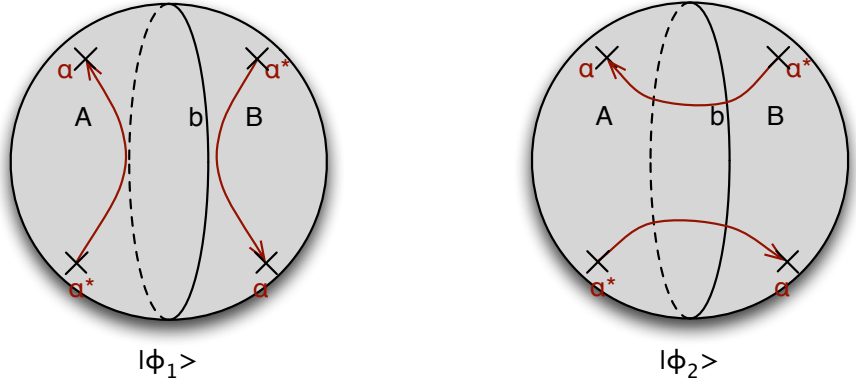}
\end{center}
\caption{Two states  represented by Wilson lines connecting punctures.}
\label{fig:caseI}
\end{figure}

If there is only one puncture in $A$, we find\cite{Dong2008} that the entanglement entropy is given by
\beq
S_A=\ln S_0^{\hat \alpha}
\eeq
On the other hand, if there are two punctures in $A$ we have two possibilities:
\begin{itemize}
\item
Case I: There is a pair of $\hat \alpha$ and $\hat \alpha^*$ in $A$ and in $B$. Each pair can fuse into the identity or into the adjoint.
For $k\geq 2$, the Hilbert space on $S^2$ with 2 pairs of $\hat\alpha$ and $\hat\alpha^*$'s is two dimensional. The entanglement entropy depends on the quantum dimensions of the conformal block.
\item
Case II: There are two $\hat \alpha$'s in $A$ and two $\hat \alpha^*$'s in $B$. The entropy now depends on which channels (representation) the quasiparticles fuse and on the choice of state (conformal block).
\end{itemize}
Let us consider first Case I. We begin with a pure state
\beq
|\phi\rangle=a|\phi_1\rangle+b|\phi_2\rangle
\eeq
where the basis states $\ket{\phi_1}$ and $\ket{\phi_2}$ are shown in \Fref{fig:caseI}.
For a mixed state with a reduced density matrix
\beq
 \rho_A=aa^*\rho_{11}+ab^*\rho_{12}+a^*b\rho_{21}+bb^*\rho_{22}
\eeq
the entanglement entropy is 
\beq
 S_{A}=\ln{{\cal S}_{0}}^{0}-\lambda_{1}\ln\lambda_{1}-(d_{\hat\alpha}^{2}-1)\lambda_{2}\ln\lambda_{2}
\eeq
where
\beq
\lambda_{1}=
\frac{|ad_{\hat\alpha}+b|^{2}}{|ad_{\hat\alpha}+b|^{2}+(d_{\hat\alpha}^{2}-1)|b|^{2}}, \quad
\lambda_{2}=
\frac{|b|^{2}}{|ad_{\hat\alpha}+b|^{2}+(d_{\hat\alpha}^{2}-1)|b|^{2}}.
\eeq

However, $\ket{\phi_1}$ and $\ket{\phi_2}$ are not orthogonal
\beq
\langle\phi_{i}|\phi_{j}\rangle={{\cal S}_{0}}^{0}d_{\hat\alpha}\left(
\begin{array}{cc}
d_{\hat\alpha}&1\\
1&d_{\hat\alpha}
\end{array}
\right).
\eeq
We can use instead an orthogonal basis
\beq
{|\phi'_{1}\rangle\choose|\phi'_{2}\rangle}=
\frac{1}{d_{\hat\alpha}\sqrt{{{\cal S}_{0}}^{0}}\sqrt{d_{\hat\alpha}^{2}-1}}\left(
\begin{array}{cc}
\sqrt{d_{\hat\alpha}^{2}-1}&0\\
-1&d_{\hat\alpha}
\end{array}
\right){|\phi_{1}\rangle\choose|\phi_{2}\rangle}
\eeq
$|\phi'_{1}\rangle$ and $|\phi'_{2}\rangle$ are conformal blocks associated with the trivial and adjoint representation $\hat\theta$, respectively, which appear in $\hat\alpha\times\hat\alpha^*$.  We have just calculated the fusion matrix. 
Here, $|\phi'_{1}\rangle$ and $|\phi'_{2}\rangle$ are the conformal blocks in one channel, $\frac{1}{d_{\hat\alpha}\sqrt{{{\cal S}_{0}}^{0}}}|\phi_{2}\rangle$ and $\frac{1}{d_{\hat\alpha}\sqrt{{{\cal S}_{0}}^{0}}\sqrt{d_{\hat\alpha}^{2}-1}}(-|\phi_{2}\rangle+d_{\hat\alpha}|\phi_{1}\rangle)$  are the blocks in the other channel.
The fusion matrix $F$ is given by
\beq
 F[\tiny{
\begin{array}{cc}
\alpha&\alpha^{*}\\
\alpha^{*}&\alpha
\end{array}
}]=
\frac{1}{d_{\hat\alpha}}
\left(
\displaystyle{
\begin{array}{cc}
1& \sqrt{\displaystyle{(d_{\hat\alpha})^{2}-1}}\\
\sqrt{\displaystyle{(d_{\hat\alpha})^{2}-1}}& -1
\end{array}
}
\right)
\eeq
$d_{\hat \alpha}$ is the quantum dimension of the fundamental representation $\hat \alpha$, and  $\sqrt{d_{\hat \alpha}^2-1}\equiv d_{\hat \theta}$, is the quantum dimension of the adjoint representation $\hat \theta$.

\begin{figure}[h!]
\begin{center}
\includegraphics[width=0.75\textwidth]{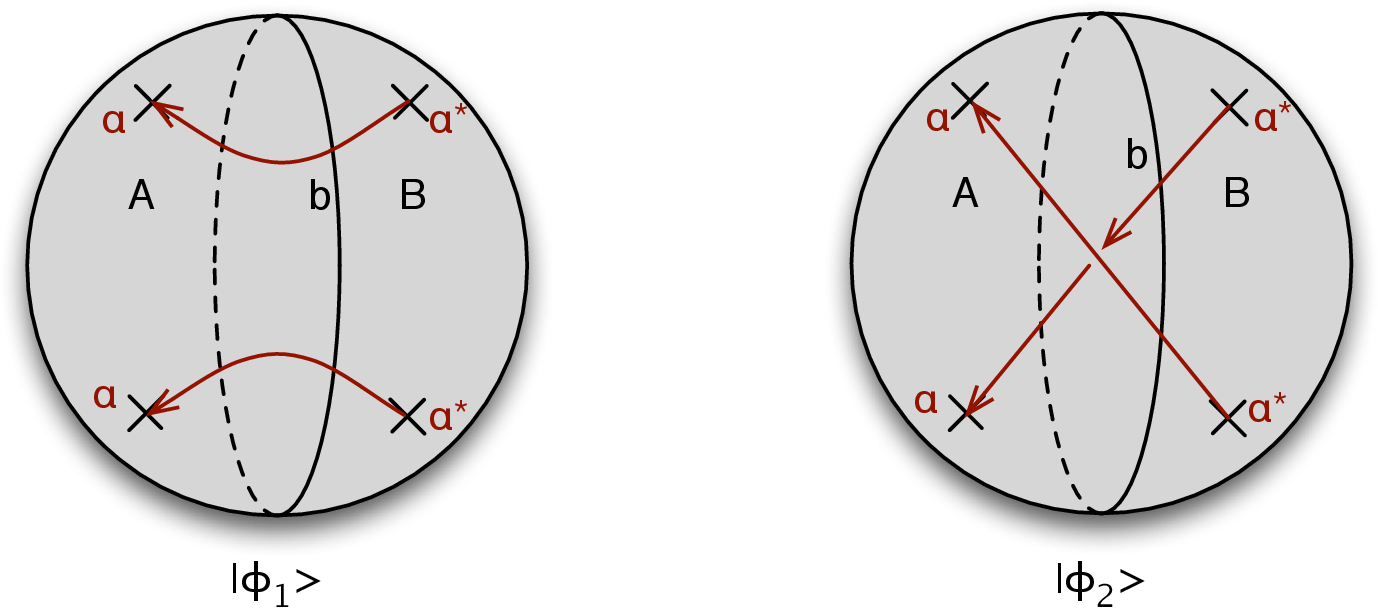}
\end{center}
\caption{}
\label{fig:caseII}
\end{figure}

Let us now turn to Case II.
We begin again with a pure state
\beq
|\phi\rangle=a|\phi_1\rangle+b|\phi_2\rangle
\eeq
with $\ket{\phi_1}$ and $\ket{\phi_2}$ depicted in \Fref{fig:caseII}.

The computation is similar to Case I except that now the Wilson loops are braided and we now fuse two fundamentals, $\alpha \time \alpha$.
The representations cut by the boundary are the symmetric ${\hat \sigma}\equiv{\hat \theta}$ and antisymmetric ${\hat \omega}\equiv 0$ rank two of two fundamentals. The entropy still has the same form with the quantum dimensions given by
\beq
d_{\hat\sigma}=\frac{[N][N+1]}{[2]},\quad d_{\hat\omega}=\frac{[N][N-1]}{[2]}
\eeq
where I use the quantum-group notation $[x]\equiv\frac{q^{x/2}-q^{-x/2}}{q^{1/2}-q^{-1/2}}$, with $q=e^{-2\pi i /(N+k)}$.
For $SU(2)_k$ the fusion matrix is the same as before.

\subsection{Entanglement Entropy of FQH states from Chern-Simons Gauge Theory}
\label{subsec:CS-FQH}

We end the discussion of the Chern-Simons theory with an application to the computation of the entanglement entropy in FQH states.

\subsubsection{Laughlin States:}
 We begin with the case of the (Abelian) Laughlin states at filling fraction $\nu=1/m$. Their effective field theory is, as we discussed earlier in this paper, a $U(1)_m$ Abelian Chern-Simons gauge theory. The results we just derived tell us that for Laughlin states the the quantum dimensions of all its $m$ quasiparticles are $d_i=1$ (as they are all Abelian). Hence, $\mathcal{S}_{00}=1/\sqrt{m}$, and the universal topological entanglement entropy is 
\beq
S_{\rm Laughlin}=-\ln \sqrt{m}
\eeq
It is straightforward to show the the entanglement entropy for {\em any} Abelian FQH state is
\beq
S_{\rm Abelian}=-\frac{1}{2} \ln g
\eeq
where $g$ is the ground state degeneracy of the Abelian FQH state on the torus.

\subsubsection{Coset $[\widehat{SU(2)/U(1)}]_k$ theories:}
\label{subsec:coset1}

Chern-Simons theory whose CFT is the coset $[\widehat{SU(2)/U(1)}]_2$ describes the two-dimensional time reversal breaking superconductors with symmetry $p_x+ip_y$. In some sense this is the simplest system with non-Abelian statistics. Here we will consider the general case of the coset $[\widehat{SU(2)/U(1)}]_k$.

In Ref.\cite{Dong2008} it is shown that the modular $\mathcal{S}$ matrix for this coset is
\beq
{	{\cal S}_{(\ell,r)}	}^{(\ell',r')}=
\sqrt{\frac{4}{k(k+2)}}		\sin\left[\frac{\pi(\ell+1)(\ell'+1)}{k+2}\right]
e^{-i\pi rr'/k}.
\eeq
 For the case of most physical interest, we have $k=2$, and  the coset primaries may be taken to be $(0;0), (1;1)$ and $(0;2)$. This is in fact just the chiral Ising model with $(0;0)\sim I$, $(1;1)\sim\sigma$ and $(0;2)\sim\psi$. The ${\cal S}$-matrix for $\left(\widehat{SU(2)/U(1)}\right)_2 $ is
\beq
{\cal S}^{\mbox{\tiny coset}}_{k=2}=\frac{1}{2}\left(
\begin{array}{ccc}
1&\sqrt{2}&1\\
\sqrt{2}&0&-\sqrt{2}\\
1&-\sqrt{2}&1\\
\end{array}
\right),
\eeq
which agrees with the results of Ref.\cite{fendley-fisher-nayak-2007c}.

Using the results we presented for Chern-Simons theory, we find that if the region of observation is simply connected, then the entanglement entropy is $S=\ln \mathcal{S}_{00}=-\ln 2$ (in this case). Our expressions earlier in this section then yield the entanglement entropy in non-trivial topologies and in the presence of quasiparticle.

\subsubsection{ Moore-Read and Read-Rezayi FQH states: pfaffian and generalized parafermion states:}
\label{subsec:coset2}

We now turn to the Moore-Read and Read-Rezayi non-Abelian FQH states, and their generalization. The filling factor of these states is $\nu=k/(Mk+2)$; $M$ even corresponds to bosonic states and $M$ odd to fermionic states \cite{Read-Rezayi-1999}. These states are described by $\widehat{[SU(2)/U(1)]_k} \times \widehat{U(1)}$ CFTs, with a suitably defined level for the $U(1)$.  Examples of these states are the well known Moore-Read pfaffian states. The the fermionic state with $k=2$ and $M=1$ has filling factor $1/2$ ($5/2$ in the experiment), and the related bosonic state at filling factor $\nu=1$ has $k=2$ and $M=0$. The states with $k>2$ are the Read-Rezayi parafermionic states. 

We will discuss both the general fermionic and bosonic states with fixed $k$ and $M$. The RCFT of interest is in all cases embedded in $\widehat{\left(SU(2)/U(1)\right)_k} \times \widehat{U(1)}_{k(Mk+2)}$. We will consider the cases of $k$ even and $k$ odd as their structure is somewhat different. Here we only present details for the simpler cases. The details of the derivations for the general case are given in Ref.\cite{Dong2008}.

By reasoning similar to the above, the resulting ${\cal S}$-matrix can be obtained by multiplying coset and $U(1)$ characters. For the pfaffian state $k=2$, the coset is a $\mathbb{Z}_2$ parafermion. The resulting $\mathcal{S}$-matrix will, up to identifications, be given by
\beq
	{{\cal S}_{(\ell,r;s)}}^{(\ell',r';s')}=
        {({\cal S}^{\mbox{\tiny coset}}_{2})_{(\ell,r)}}^{(\ell',r')}
	{(\mathcal{S}^{U(1)_{4M+4}} )_s}^{s'}.
\eeq
Primaries of this theory will be given by products of the $\mathbb{Z}_2$ primaries $\{ I,\sigma,\psi\}$ with $U(1)_{4M+4}$ primaries of the form ${\cal O}_{\ell/p}$. We seek a set of such operators that close under operator products and are local with respect to a suitable extended current algebra, which will be generated by $J_\pm \sim\psi \; e^{\pm i\sqrt{M+1}\phi}$, where $\psi$ is the Majorana fermion of $\mathbb{Z}_2$. For simplicity, we will consider two cases here, $M=0$ (take $p=2$, $p'=1$, radius $R=1$) and $M=1$ (take $p=4$, $p'=1$, radius $R=\sqrt{1/2}$).

In the case of $k=2$ and $M=0$, we find the integer-weight $J_\pm \sim \psi \; e^{\pm i \phi}$ as suitable extended currents. Requiring locality of operator products, we then find that the primaries of this theory are given by $I, \psi,\sigma e^{i\phi/2}$ (all others are related to these by action of $J_\pm$). These in fact are just the primaries of $\widehat{SU(2)}_2$, as we should expect. This is the bosonic pfaffian state. 

In the case of $k=2$ and $M=1$ (the fermionic pfaffian state), we find $J_\pm \sim \psi \; e^{\pm i\sqrt{2} \phi}$ as suitable extended currents\cite{moore-read-1991}. Requiring locality of operator products, we then find that the primaries of this theory are given by 
\beq\label{fpfstates}
I,\psi,\sigma e^{\pm i\phi/2\sqrt{2}},e^{\pm i\phi/\sqrt{2}}.
\eeq
This set closes under fusion (up to the action of $J_\pm$). These operators have weights\footnote{The notation $(\ell,r;s)$ represent the coset weights $(\ell,r)$ and the $U(1)$-charge $s$.} $(0,0;0)$, $(0,2;0)$, $(1,1;\pm1)$ and $(0,0;\pm2)$ respectively. We can then read off the ${\cal S}$-matrix:
\beq
{\cal S}=
\frac{1}{2\sqrt{2}}\left(
\begin{array}{cccccc}
1&1&\sqrt{2}&\sqrt{2}&1&1\\
1&1&-\sqrt{2}&-\sqrt{2}&1&1\\
\sqrt{2}&-\sqrt{2}&0&0&+i\sqrt{2}&-i\sqrt{2}\\
\sqrt{2}&-\sqrt{2}&0&0&-i\sqrt{2}&+i\sqrt{2}\\
1&1&i\sqrt{2}&-i\sqrt{2}&-1&-1\\
1&1&-i\sqrt{2}&+i\sqrt{2}&-1&-1
\end{array}
\right),
\label{eq:Smatrix-pfaffian}
\eeq
from which one can read-off the total quantum dimension is ${\cal D}=2\sqrt{2}$. 

We will now consider the interesting example of the parafermionic states at $k=3$ and $M=1$: the Read-Rezayi parafermionic state for fermions at filling factor $2+2/5$. The $k=3$ coset has primaries at
$(\ell,r) = (0,0),\ (1,\pm 1),\ (2,0),\ (3,\pm 1)$, 
which we will refer to as $I,\sigma_{\pm}, \epsilon, \psi_\pm$ respectively.
Explicitly, denoting $s_p\equiv \sin (\pi p/5)$, we have
\beqn
{\cal S}^{\mbox{\tiny coset}}_{k=3}
&=&\frac{2}{\sqrt{15}}\left(
\begin{array}{cccccc}
s_1&s_2&s_2&s_2&s_1&s_1\\
s_2&e^{-i\pi/3}s_1&e^{+i\pi/3}s_1&-s_1&-e^{-i\pi/3}s_2&-e^{+i\pi/3}s_2\\
s_2&e^{+i\pi/3}s_1&e^{-i\pi/3}s_1&-s_1&-e^{+i\pi/3}s_2&-e^{-i\pi/3}s_2\\
s_2&-s_1&-s_1&-s_1&s_{2}&s_{2}\\
s_1&-e^{-i\pi/3}s_2&-e^{+i\pi/3}s_2&s_{2}&-e^{-i\pi/3}s_1&-e^{+i\pi/3}s_1\\
s_1&-e^{+i\pi/3}s_2&-e^{-i\pi/3}s_2&s_{2}&-e^{+i\pi/3}s_1&-e^{-i\pi/3}s_1\\
\end{array}
\right).
\nonumber \\
&&
\eeqn
For this case there is an extended algebra generated by the $h=3/2$ operator $Q_+=\psi_+ \; e^{5i\phi/\sqrt{15}}$, where $\phi$ is a free boson of the $U(1)$ theory that we are attaching. Representative primaries are
$(\ell,r;s) = (0,0;0)$, $ (3,-1;1)$, $ (3,1;2)$, $ (0,0;3)$, $ (3,-1;4)$ and 
$(\ell,r;s) =  (2,0;0)$, $ (1,-1;1)$, $ (1,1;2)$, $ (2,0;3)$, $ (1,-1;4)$. 
One can check that these have local OPE's with $Q_+$ and are closed under fusion. As we will see, it is convenient to group them into groups of $k+2=5$, as given. The theory obtained this way is actually an $N=2$ superconformal theory, with supercharges $Q_\pm$ ($Q_-$ being $\psi_-\; e^{-5i\phi/\sqrt{15}}$). $Q_+$ groups collections of conformal primaries together, {\it i.e.\/}, 
$\{(0,0;0), (3,1;5), (3,-1;10)\}$, 
$\{(3,-1;1), (0,0;6), (3,1;11)\}$, 
$\{(3,1;2), (3,-1;7), (0,0;12)\}$,  
\\
$\{(0,0;3), (3,1;8), (3,-1;13)\}$,
and   
$\{(3,-1;4), (0,0;9), (3,1;14)\}$ 
and 
$\{(2,0;0), (1,1;5), (1,-1;10)\} $, 
\\
$\{(1,-1;1), (2,0;6), (1,1;11)\} $, ~~
$\{(1,1;2), (1,-1;7), (2,0;12)\} $,  
$\{(2,0;3), (1,1;8), (1,-1;13)\} $,  
and 
$\{(1,-1;4), (0,0;9), (1,1;14)\} $.
Each of these triplets represents a superconformal family.
When we compute the ${\cal S}$-matrix with respect to the extended symmetry, we treat these groupings as one. That is, computing the $\mathcal{S}$-matrix element on the grouping gives a $3\times 3$ identity matrix times a factor. We collect those factors into the following $\mathcal{S}$-matrix.
\beqn\label{k3M1}
{\cal S}^{FRR}_{k=3}&=&\frac{2}{5}\left(
\begin{array}{cc}
\sin (\pi /5)&\sin (2\pi /5)\\
\sin (2\pi /5)&-\sin (\pi /5)\\
\end{array}\right)
\otimes
\left(\begin{array}{ccccc}
1&1&1&1&1\\
1&\omega_2&\omega_4&\!\!\!\!\!\! \omega_1&\omega_3\\
1&\omega_4&\omega_3&\omega_2&\omega_1\\
1&\omega_1&\omega_2&\omega_3&\omega_4\\
1&\omega_3&\omega_1&\omega_4&\omega_2\\
\end{array}\right).
\eeqn
where we have used the $U(1)$ ${\cal S}$-matrix is ${{\cal S}_{s}}^{s'}=\frac{1}{\sqrt{15}}e^{2\pi i ss'/15}$.
Above we used the notation is $\omega_{p}=e^{2\pi i p/5}$. 
The coefficient out front is $\frac{2}{\sqrt{15}}\cdot \frac{1}{\sqrt{15}}\cdot \frac{15}{5}$, the factors being the coefficients of the coset $\mathcal{S}$-matrix, the $U(1)$ $\mathcal{S}$-matrix and the order of the automorphism (5 in 15), respectively.
Note that it is easy to read off then the total quantum dimension 
\beq
{\cal D}=\frac{1}{{{\cal S}_0}^0}=\frac{5}{2\sin(\pi/5)}=
{\cal D}=\sqrt{5 + 5(s_2/s_1)^2}=\sqrt{5 (1+\phi^2)},
\eeq
where here $\phi=(\sqrt{5}+1)/2$ denotes the Golden Ratio.

In Ref.\cite{Dong2008} it is shown that  for  general $k$ and $M$, the primaries are the highest weight states of the form
\beq
\psi_{(\ell,\ell-2[\frac{n}{M}])} \;
\exp \left(i\frac{\ell+nk-(Mk+2)[\frac{n}{M}]}{\sqrt{k(Mk+2)}}\phi\right) 
\eeq
or
\beq
\psi_{(\ell,\ell-2[\frac{n-1}{M}])} \;
\exp \left(i\frac{\ell+nk-(Mk+2)[\frac{n-1}{M}]}{\sqrt{k(Mk+2)}}\phi \right),
\eeq
where $\psi_{(\ell,r)}$ are  $\mathbb{Z}_k$-parafermion primaries, $n$ and $\ell$ are integers , $[x]$ is the closest integer to $x$. For general $k$ the $\mathcal{S}$-matrix is given by
\beqn
&&\!\!\!\!\!\!\!\!\!\!\!\! \!\!\!\!\!\! \!\!\!\!\!\!\!\!\!\!\!\! \!\!\!\!\!\!   {\mathcal{S}_{\{\ell;n\}}}^{\{\ell';n'\}}=\frac{2}{\sqrt{(k+2)(Mk+2)}}\sin \left[\frac{\pi(\ell+1)(\ell'+1)}{k+2}\right] \; \exp \left(\frac{\pi i(-M\ell\ell'+2\ell n'+2\ell' n+2knn')}{Mk+2}\right).\nonumber \\
&&
\eeqn
One can read off from this the total quantum dimension $\mathcal{D}$ for all $M$ and $k$, since
\beq
\frac{1}{\mathcal{D}}={{\cal S}_0}^0=\frac{2}{\sqrt{(k+2)(Mk+2)}} \; \sin\left(\frac{\pi}{k+2}\right).
\eeq
from where the value of the topological entanglement entropy for a simply connected region can be obtained.

\section{Outlook}
\label{sec:outlook}

We discussed the behavior of the entanglement entropy near quantum phase transitions and in topological phases. In some special cases, the entanglement entropy of 2D QCPs with conformally invariant wave functions has a
universal logarithmic terms. However, if the logarithmic term is absent the $O(1)$ term is universal.
We also showed that in a topological phase the finite term in the entanglement entropy is a universal property of the phase, which in general is different from its value at the QCP.

We computed the topological entanglement entropy for Chern-Simons gauge theories.
The entanglement entropy of abelian and non-abelian FQH states is given in terms of the modular $\mathcal{S}$ matrix for the effective CS theory.
This requires to glue $U(1)$ charge sector and the coset $\left(SU(2)/U(1)\right)_k$ neutral sector.

For a simply connected region it is universal and depends only on the total quantum dimension.
For regions which are not simply connected, the entropy is additive. The entropy of disjoint regions on a torus depends on the effective quantum dimension and on the state on the torus.
The entropy for a simply connected region on the sphere with 4 quasiparticles (punctures) depends on the conformal block.
The entropy depends on the quantum dimensions and on the particular state that is chosen. The change of basis depends on the fusion matrix and on the conformal weights as well.

These results suggest that it may be possible to determine the structure of the topological field theory by means of entanglement entropy measurements.

We close with some comments on the question on how (or even whether) the entanglement entropy may be measured experimentally. Our results indicate that in the case of the conformal quantum critical points in 2D there is an intimate relationship between the entanglement entropy and the Affleck-Ludwig vacuum degeneracy of the associated Euclidean CFT. On the other hand the work by Fendley, Fisher and Nayak on entanglement at point contacts of the non-Abelian FQH states\cite{Fendley2007c} also relates the entanglement entropy (change) of the bulk topological fluid, to the (change) of the Affleck-Ludwig entropy of the point contact, a property of the edge states.  These seemingly unrelated results suggest that there may be a deeper connection. More important perhaps changes in the bulk entanglement entropy can be measured by monitoring properties of a suitably defined point contact. Recent work by Klich and Levitov\cite{klich-2009} in the context of a simple (and hence solvable) system of free fermions, suggest that this may be possible. We have preliminary results that support this idea in a more general context\cite{hsu-2009b}.

\ack
This review paper was prepared borrowing heavily from the papers I wrote with my collaborators Joel E. Moore, Chetan Nayak, Kareljan Schoutens, Eddy Ardonne, Paul Fendley, Stefanos Papanikolaou, Benjamin Hsu, Eun-Ah Kim, Michael Mulligan, Eytan Grosfeld,  Shiying Dong, Robert G. Leigh and Sean Nowling. I am grateful to them for sharing their insights with me and for explaining me many things. This work was supported in part by the National Science Foundation through the grant DMR 0758462.

\section*{References}


\providecommand{\newblock}{}

\end{document}